\documentclass[reprint,superscriptaddress,amsmath,amssymb,aps,pre]{revtex4-2}

\usepackage{amsmath,amssymb,graphicx,enumerate,bm,dcolumn}
\usepackage{balance}
\usepackage{hyperref}
\usepackage{float}
\usepackage[english]{babel}
\usepackage{array}

\begin{document}
%\preprint{APS/123-QED}

%%%TITLE, AUTHORS AND ABSTRACT%%%
\title{The pseudo-two-dimensional dynamics in a system of macroscopic rolling spheres}

\author{M. A. L\'opez-Casta\~no}
    \email{malopez00@unex.es}
    \affiliation{Departamento de F\'isica, Universidad de Extremadura, Avda. Elvas s/n, 06071 Badajoz, Spain}
\author{J. F. Gonz\'alez-Saavedra}
    \affiliation{Departamento de F\'isica, Universidad de Extremadura, Avda. Elvas s/n, 06071 Badajoz, Spain}
\author{A. Rodr\'iguez Rivas}
    \affiliation{Departamento Matem\'atica Aplicada II, Escuela Polit\'ecnica Superior, Universidad de Sevilla, \\ Virgen de Africa, 7, 41011 Seville, Spain}
\author{E. Abad}
    \affiliation{Departamento de Física Aplicada, Centro Universitario de M\'erida, Universidad de Extremadura, \\ 06800 M\'erida, Spain}
    \affiliation{Instituto de Computaci\'on Cient\'ifica Avanzada (ICCAEx), Universidad de Extremadura, \\ Avda. Elvas s/n, 06071 Badajoz, Spain}
\author{S. B. Yuste}
    \affiliation{Departamento de F\'isica, Universidad de Extremadura, Avda. Elvas s/n, 06071 Badajoz, Spain}
    \affiliation{Instituto de Computaci\'on Cient\'ifica Avanzada (ICCAEx), Universidad de Extremadura, \\ Avda. Elvas s/n, 06071 Badajoz, Spain}
\author{F. {Vega Reyes}}
    \email{fvega@unex.es}
    \affiliation{Departamento de F\'isica, Universidad de Extremadura, Avda. Elvas s/n, 06071 Badajoz, Spain}
    \affiliation{Instituto de Computaci\'on Cient\'ifica Avanzada (ICCAEx), Universidad de Extremadura, \\ Avda. Elvas s/n, 06071 Badajoz, Spain}

\date{\today}

\begin{abstract}
    We study in this work the dynamics of a collection of identical hollow spheres
    (ping-pong balls) that rest on a horizontal metallic grid. Fluidization is
    achieved by means of a turbulent air current coming from below. The upflow is
    adjusted so that the particles do not levitate over the grid, resulting in
    quasi-two-dimensional dynamics. We show the behavior of diffusion and
    correlations in this system is
    particularly rich. Noticeably as well (and related to the complex dynamical
    behavior), a variety of phases appear, with important peculiarities with respect
    to analogous set-ups. We observe gas, liquid, glass and hexagonal crystal
    phases. Most notably, we show that the melting of the hexagonal crystal occurs
    in co-existence with a liquid phase. This strikingly differs from the
    corresponding transition in a purely two-dimensional systems of air-fluidized
    disks, for which no phase coexistence has been reported in the bibliography. 
\end{abstract}
%%%END OF TITLE, AUTHORS AND ABSTRACT%%%

\maketitle

%%%MAIN TEXT%%%%
\section{Introduction}
    
The dynamics of macroscopic particle systems has attracted the interest of
physicists and engineers since long ago \cite{F31,BN47,B38}. Its significance is
partly due to the potential analogies with microscopic particle systems, and partly
because the manipulation of granular materials finds widespread application in
industry \cite{JNB96}. Therefore, the understanding of what has been termed `granular
dynamics' is important both from the point of view of theory and applications. More
specifically, there has been an increasing interest on two-dimensional (granular
systems over the last decades \cite{OU98,Melby2005}. 

Granular media share important similarities with molecular matter (as already outlined by O. Reynolds in 1885 \cite{R85}), but yet there are also significant differences and peculiarities.
%The existence of these differences triggered and inspired research
%on granular matter, with focus on phenomena that had
%already been known to exist in molecular matter, but now paying special
%attention to the peculiarities in granular matter.
Convection and turbulence \cite{Eshuis2007,Khain2003}, jamming \cite{LN98}, Brownian motion \cite{OLDLD04}, crystallization \cite{RIS06,CMS12,OU98}, and other phenomena well known in molecular matter have also been observed in granular matter, but they are usually more complex and they often exhibit peculiarities. Furthermore, some of the phenomenology reported in previous works is exclusive to granular matter, such as inelastic collapse \cite{OU98} and clustering instabilities \cite{GZ93}.

In particular, the attention drawn by crystallization and ordering phenomena in 2D
granular systems is partly due to the impact of 2D equilibrium theories in the field
of condensed matter \cite{KT73,KT16}. The seminal work by Kosterlitz, Thouless,
Halperin, Nelson, Young \cite{KT73,NH79,Y79} (subsequently extended by others
\cite{S88}) highlights the role of spatial dimension, as it predicts fundamental
differences in the behavior of two-dimensional (2D) systems with respect to that of
their three-dimensional counterparts.
For instance, theoretical findings and experimental observations \cite{KT73,NH79,Y79}
show that the crystal melting transition in 2D equilibrium systems is in general
continuous and defect-mediated \cite{F64}. The explanation to this 2D transition is
usually referred to as the Kosterlitz-Thouless-Halperin-Nelson-Young (KTHNY) theory \cite{S88,OU05}.
This emphasizes the interest of studying 2D granular systems. An additional advantage
of such systems is that both the experimental measurements and the characterization
of many properties of interest are often more straightforward than in 3D systems \cite{RRSS18}.

In order to induce granular matter thermalization, some kind of energy input is necessary, since energy is lost in macroscopic particle collisions \cite{G03}. Depending on the type of driving, experimental work in 2D systems has relied mostly on air fluidization \cite{Oger1996,MBF81,OLDLD04} or shaking, either tangent \cite{KHTPB03} or perpendicular to the plane to which the motion is constrained. Up to some exceptions \cite{Pontuale2016,Sal18}, in most works the plane in which the particle motion takes place coincides with the horizontal plane; hence, tangent and perpendicular shaking are equivalent to horizontal and vertical shaking, respectively.
Additionally, there are some interesting shaking experiments with no gravity \cite{Grasselli2015} (for which the horizontal direction is of course not defined).
However, more recent work makes use of alternative methods of thermalization with the advantage that boundary friction with boundaries effects are not present, such as an AC electric field on charged particles \cite{GB17} or acoustic
levitators \cite{LSVJ19}.

For the purpose of studying phase transitions, horizontal shaking experiments differ in that, since field gradients are generated from the boundaries, particles located near the walls will experience a net injection of energy while particles in the bulk will suffer mainly dissipative collisions, thereby giving rise to inherently inhomogeneous systems \cite{PM13}; which renders the analysis of order transitions more difficult. In vertical shaking (quasi) 2D experiments, however, homogeneous dynamics can more easily be achieved. A variety of very interesting results have been obtained in vertical shaking experiments with spheres \cite{PMKW78,OU98,Melby2005,RIS06,PM11,CMS12,MS16,SK19,CMSSGS19}. In particular the existence of a liquid-to-crystal continuous transition mediated by the \textit{hexatic} phase has been confirmed,
in agreement with the predictions of the KTHNY theory for equilibrium systems \cite{OU05,KT15}.

In air tables, an appropriately adjusted air current flowing from below prevents
levitation of the particles (the dynamics is thus restrained to a single plane), and
also generates thermal-like motion via the stochastically fluctuating turbulent
wakes that are caused by the interstitial air upflow \cite{OLDLD04}. Moreover, the
dynamics is found to be homogeneous if the upflow is homogeneous as
well \cite{MBF81}. In this way, horizontal dynamics is effectively achieved  (i.e.,
no translational kinetic energy is stored in the vertical degree of freedom) for
both plane (disks \cite{MBF81,Oger1996}) and non-plane particles (spheres
\cite{OLDLD04}).

At this stage, a comment on a subtle yet important difference between air table
experiments and vertical shaking experiments \cite{Melby2005} is in order. In the
latter, there is an intrinsic (non-measurable) fraction of the translational
kinetic 
energy directly input in the vertical direction via mechanical collisions between
the particles and the shaking boundaries \cite{Melby2005}. However, in air tables the
motion of spheres outside the horizontal plane is limited to sphere rolling,
implying that there are no vertical displacements of the center of mass of the
particles. For the sake of precision, we will make use of the term \textit{quasi-2D} 
or \textit{pseudo-2D} to refer to the dynamics of rolling spheres described in this
work (as already explained, for an analogous but slightly different reason, vibrated
thin layers \cite{Melby2005} are also referred to as \textit{quasi-2D} systems) \cite{MS16} .

It is also important to note that, according to the type of particles in air tables, we can distinguish
between works dealing with flat particles (disks, usually \cite{Oger1996,Grasselli2015}), to which
we will refer as two-dimensional (2D) systems and works dealing
with non-plane particles, most notably, spheres \cite{OLDLD04} (as we said, we will
refer to these systems as being quasi-2D). Thus, in our work, we are specifically interested in pseudo-2D dynamics, and not in strictly 2D dynamics.

As a lead-in to relevant results found throughout this work: we have carried out a
preliminary description of phase behavior (most notably, crystallization processes)
of rolling spheres, which had not been addressed in previous works. We will see that
set of spheres on a horizontal air table may undergo a variety of different phases,
ranging from the low density granular gas to highly packed crystals; unlike in
quasi-2D vertical shaking  experiments, where low density phases are not observable
in wide regions of the parameter space \cite{NRTMS14}. Additionally, we find that
repulsion forces between the spheres (of hydrodynamic origin \cite{Ojha2005}) are at
play in our system, and this will have a crucial impact on the phase behavior. We
have included a brief quantitative description of  these phases through the
computation of the appropriate bond-orientational order parameter and Voronoi
diagrams.  
We also report results on the velocity distribution function (investigating the
causes of deviations from a purely Maxwellian behavior), velocity autocorrelation and
radial distribution function, these are important to describe the mechanisms by which
particles interact with each other and to characterize the observed phases. 
Besides that, we have also studied the diffusive nature of our system, which is an
aspect often overlook in previous works on similar systems. An interesting take-out
is the finding of some regions (in the density-temperature parameter space) where the
observed behavior is markedly subdiffusive (this being associated in some cases with
a glass-like phase). We also encourage the reader to take a look at the supplementary
material \cite{suppl} where we have included a novel result regarding the
non-monotonous behavior of the granular temperature and some illustrative movies. 
 
This paper is organized as follows. In the next section, we describe the experimental
system and the particle tracking methods \cite{CG96} we have used.
%We also analyze the inherent sources of measurement error (EL ANALISIS DE ERROR SE HA PASADO AL MATERIAL SUPLEMENTARIO, ¿NO?), and the
%homogeneity of the upflow and the generated particle dynamics.
In Section \ref{dynamics} we analyze the behavior of dynamical properties (distribution function, velocity autocorrelation and diffusion). Section \ref{structural} discusses the ordering properties of the system
and the emergence of phase transitions.
In Section \ref{conclusions} we discuss the results and outline some open problems that could be studied with similar experimental set-ups.

\section{Description of the system}
\label{ExpM}

We perform experiments with a variable number $N$ of identical spherical particles. Specifically, our particles are ping-pong balls with diameter  $\sigma = 4~\mathrm{cm}$ (ARTENGO\texttrademark~brand balls, made of ABS plastic with mass density $0.08~\mathrm{g\,cm}^{-3}$). The spheres rest on a metallic mesh (circular holes of $3~\mathrm{mm}$ diameter arranged in a triangular lattice) and are enclosed by a circular wall made of
polylactic acid (PLA). The diameter of this circular boundary is $D=72.5~\mathrm{cm}$ and its height is $h\simeq 4.5~\mathrm{cm}>\sigma$. Thus, the
total area of the system available to the spheres is $A=0.413~\mathrm{
  m}^{2}= 328.65\times\pi(\sigma/2)^2$, which means that up to $N_\mathrm{max}=(\pi/\sqrt{12})\times 328.65\simeq298$ balls can fit in our system neglecting boundary effects (the $\pi/\sqrt{12}\simeq 0.9069$  factor corresponds to the maximum packing fraction for disks in an infinite system \cite{B83}).

\begin{figure}[!t]
    \centering
        (a)\hspace*{0.75\columnwidth}~\\
        \includegraphics[height=5.25cm]{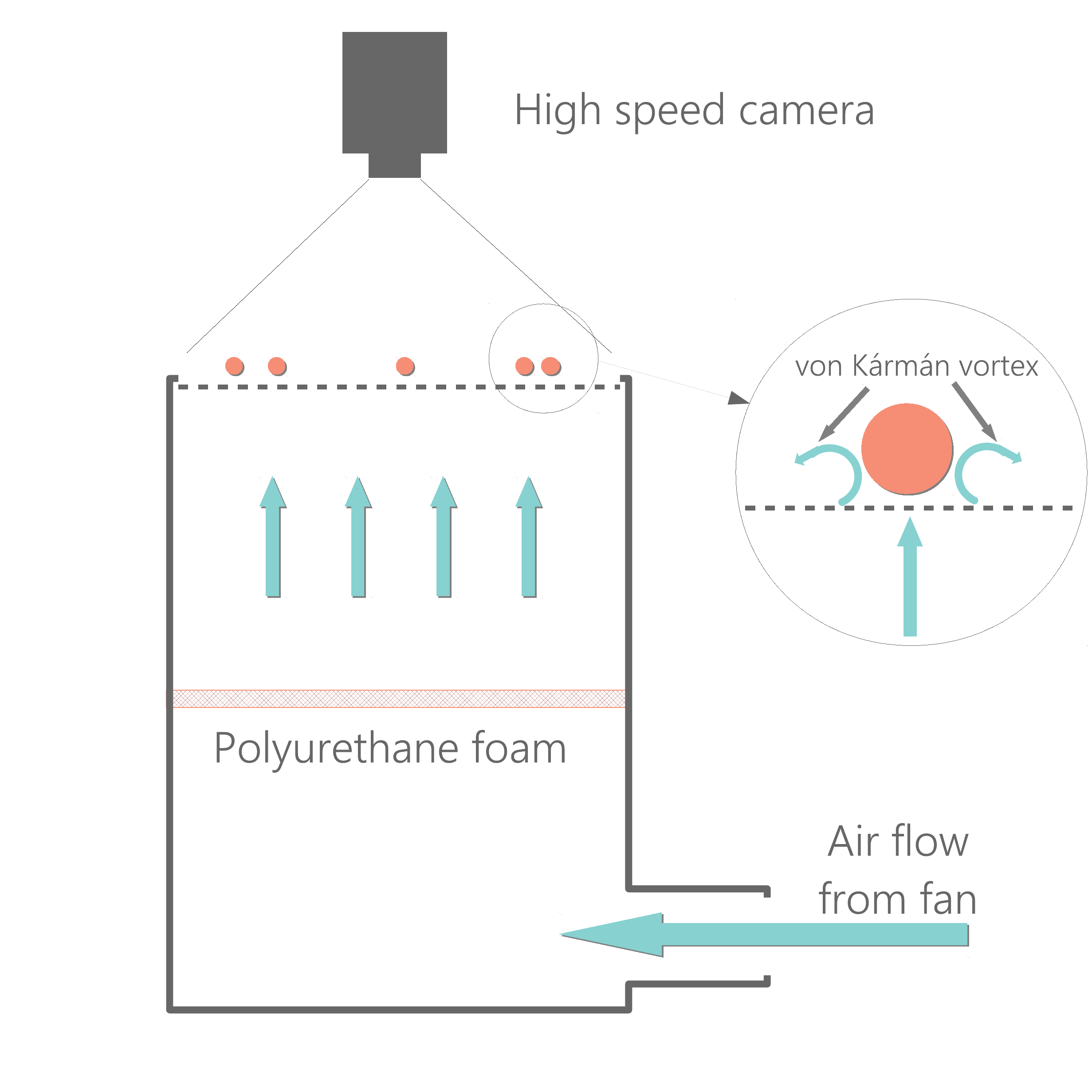}\hspace{1cm}\\
        (b)\hspace*{0.75\columnwidth}~\\
       \includegraphics[height=4.25cm]{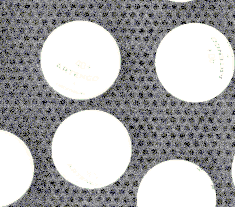}
    \caption{(a) Sketch of the experimental set-up. (b) Sample image showing the relative size of balls and grid holes.}
    \label{scheme}
\end{figure}

%This surface in which particles are is mounted on a box ($ 90\mathrm{x}90\mathrm{x}120 $ cm).
A state of the particulate system with stationary statistical properties is achieved
by means of a vertical air flow in upward direction, as depicted in Figure
\ref{scheme}. This upflow through the metallic grid generated with a fan,
SODECA\texttrademark~$\mathrm{HCT-71-6T-0.75/PL}$, and has stream velocities in the range
$u_{air}=[$2 - 5.5$]$ $\mathrm{m/s}$. We have observed an approximately linear
relationship between  $u_{air}$ and fan power. An intermediate foam ($\sim
2~\mathrm{cm}$ thick) homogenizes the air current from the fan.

In order to assess the homogeneity of the flow throughout all the interstitial
regions of the system, the air flow distribution over the grid was measured with a
turbine digital anemometer plugged to a computer for the sake of data collection. We
took measurements over a square grid of regularly spaced points on the table, and
found local deviations of the air current of less than $10\%$ with respect to the
average $u_{air}$. The air current coming from the fan produces turbulent wakes as it
flows past the spheres \cite{vD82}. We thus achieve a pseudo-two-dimensional
particle dynamics, since the relevant particle motion is restrained to the grid plane
(for more details on our particle fluidization mediated by turbulent air flow, see
the Supplementary material \cite{suppl}).

Summarizing, our experimental system has the following properties: 1) It is a
many-particle system; 2) energy input (in absence of particles) can be measured  and
is found to be homogeneous; 3) motion is contained in a horizontal plane (the grid),
and as a consequence gravity does not single out a predominant direction for
in-plane particle movement. %4)
                                %symmetry-break is observed in the form of a
                                %hexagonal arrangement of the particles.

A series of experiments has been carried out by modifying the values of air flow
intensity ($2~\mathrm{m/s}\leq u_{air} \leq 5.5~\mathrm{m/s}$) and packing
fraction $\phi\equiv N(\sigma / D)^{2}$, ($0.03\leq \phi \leq 0.79$). We have recorded a $99.92~\mathrm{s}$ clip of each
experiment with a high-speed camera (Phantom\texttrademark~ VEO 410L) at
$250~\mathrm{frames/s}$, or fps, (well below the maximum operational speed of
our camera); i.e. the camera records a new image every $\Delta
t_\mathrm{fps}=1/250~\mathrm{s}$. Particle positions are detected and tracked
throughout the movies by means of a particle-tracking algorithm \cite{CG96,opencv} that,
after adjusting for our particles and illumination conditions, is
applied to the digital images taken by the camera. Images are recorded at the camera
maximum working resolution (1200 $\times$ 800 pixels). In order to obtain high
resolution images of the spheres (with 80 pixels per particle diameter), the camera
was zoomed on the central region of
the system; i.e., highly accurate particle position and velocity
measurements were taken. More details on particle-tracking and experimental methods
and as well as the measurement accuracy we achieved can be found
in the supplementary material \cite{suppl}.

\section{Dynamical properties}
\label{dynamics}

Air-fluidized granular 2D or pseudo-2D systems have already been studied by other
researchers. The closest analogs to our system may be found in the works involving
air table experiments with disks (2D dynamics)
\cite{Lemaitre1991,Lemaitre1993,Ippolito1995,Oger1996} and with spheres (pseudo-2D dynamics) \cite{OLDLD04,Ojha2005,Abate2005} In the system with spheres, several series of experiments were
initially performed with a single ping-pong ball \cite{OLDLD04} and a small number
of them \cite{Abate2005,Ojha2005}, in order to characterize microscopic fluctuations and
particle-particle and wall-particle forces. It was only later that,
experiments were performed with larger sets of spheres in order to study jamming
conditions \cite{Abate2006}. 

Inspired by these previous works, in what follows we will extend previous studies by
providing a comprehensive description of the different dynamic properties displayed
by a system with a relatively large number of spheres. A full exploration of the
parameter space defined by the packing fraction $\phi$ and the granular
temperature $T$ can be achieved by controlling the number of particles $N$ and the air
upflow velocity $u_\mathrm{air}$. We must also note that some aspects of our system
dynamics differ from those of previous works for closely related systems; in particular, in our
experiments particles do not appear to be trapped in a harmonic potential, as
opposed to previous results \cite{OLDLD04}. Furthermore, in contrast with some
previous results \cite{Oger1996,Abate2008}, we find that granular 
temperature does not decrease monotonically with particle density.
These differences will be further discussed in the remainder of this paper.

\par\bigskip

\subsection{Distribution function and velocity autocorrelation}
\label{fv}

% Once the particle trajectories have been tracked, we are able to produce a
% statistical analysis of the dynamics of the system.

\begin{figure}[!t]
  \centering
    (a) \hspace*{0.75\columnwidth}~\\
    \includegraphics[width= 0.855 \columnwidth]{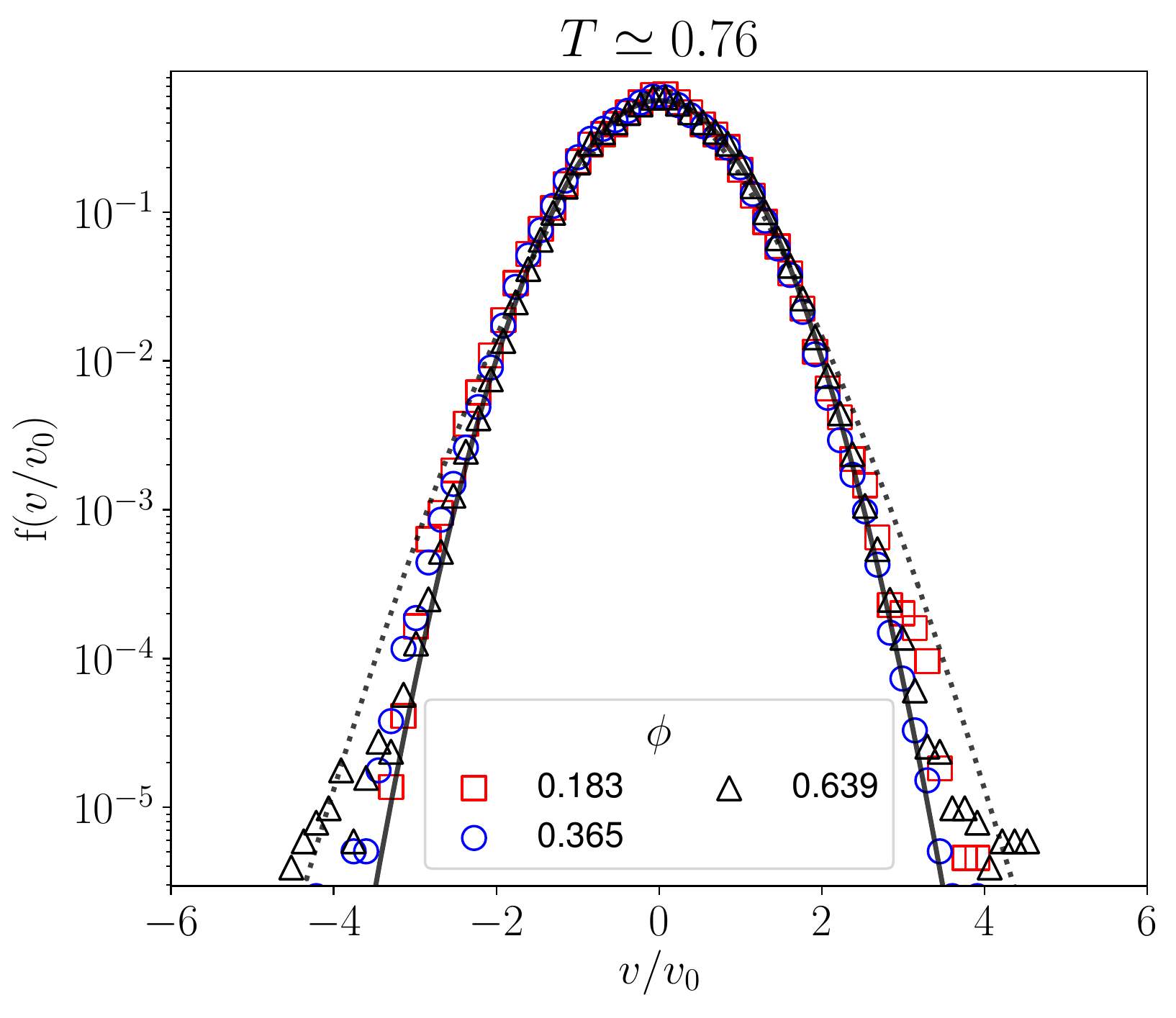} \\
    (b) \hspace*{0.75\columnwidth}~\\
    \includegraphics[width=0.85 \columnwidth]{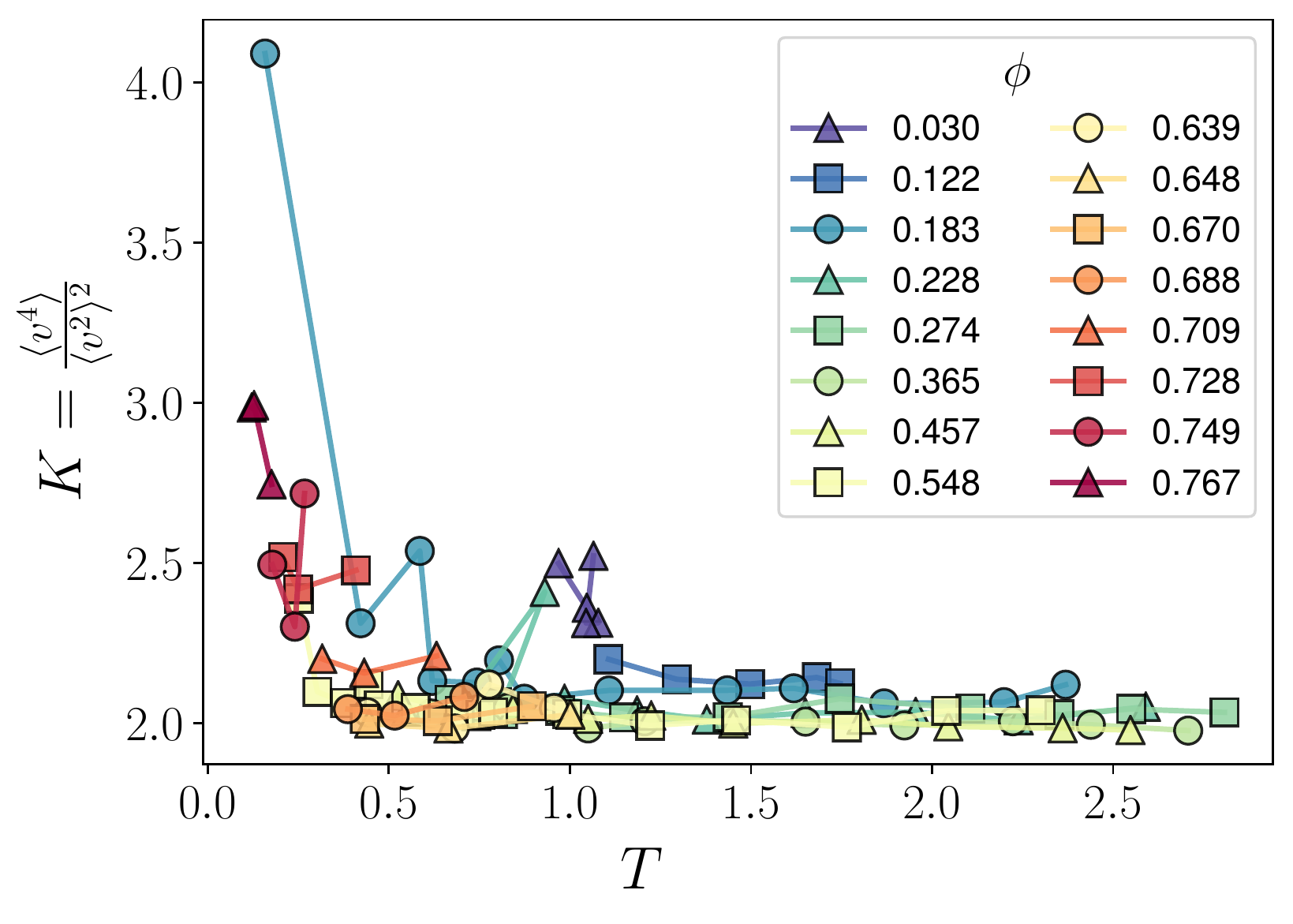}
    \caption{(a) Velocity distribution functions in logarithmic
      scale for a  series taken  at approximately constant
      granular temperature. The experimental data reveal that high density systems with exhibit more pronounced non-Maxwellian high-energy tails at $T=0.76$.  (b) Here we represent the kurtosis for constant density
      series vs. $T$. }
    \label{log_fv}
\end{figure}

In Figure~\ref{log_fv} (a) we show the distribution function $f(c)$ of the rescaled
velocity $c\equiv v/v_0$ (with $v_0\equiv(2T/m)^{1/2}$ being the thermal velocity and
$T\equiv (m/2)\langle v^2\rangle$ the granular temperature, and $\langle \cdots
\rangle$ denotes ensemble average). 
Except when specified otherwise, magnitudes are dimensionless. We use particle diameter
$\sigma$, seconds $s$ and particle mass  $m$ as units for length,
time and mass respectively.

The results show a clear tendency to deviate from
the Maxwellian distribution function (represented here by a solid line); this trend
being stronger the denser the system. As observed in previous experimental works on
quasi-2D granular dynamics, as the tails deviate from the Maxwellian, they become
exponential-like \cite{Prevost2002,OU99,Scholz2017}.

Moreover, it is interesting to note that this behavior was previously reported for
constant particle density series with increasing granular temperature, but not
for (approximately) constant temperature series, as displayed in Figure~\ref{log_fv}~(a).
We chose to compare systems with similar temperature in order to isolate the effects of modifying $\phi$ from the effects of changing the energy input-dissipation balance.

There is a certain difficulty in creating these constant temperature series, since the range of attainable granular temperatures can be very narrow depending on particle density. % In
% order to illustrate this point, the
% granular temperature value chosen for Figure~\ref{log_fv}~(a) has been marked in
% Figure~\ref{temperature} with a horizontal dashed line.

Figure~\ref{log_fv} (b) shows the kurtosis $K=\langle v^4\rangle/\langle v^2\rangle^2$ of the distribution function, which can be used to quantify the deviations from the Maxwellian distribution. As we can see, there
is also a strong overall tendency to deviate significantly from the Maxwellian at
low temperatures and low densities (see Figure~\ref{log_fv} b). This probably signals the prevalence in this regime of friction effects due to the interaction between the irregular mesh surface and the balls, and as we will see later, can also be an indication of ordering processes.

\begin{figure*}
  \centering
  \begin{tabular}{ c c }
%\hline
%\includegraphics[width=.46 \textwidth]{./figs/Fig6a.pdf}
    (a) \hspace*{0.25\textwidth} & (b) \hspace*{0.25\textwidth} \\
    \includegraphics[width= 0.3\textwidth]{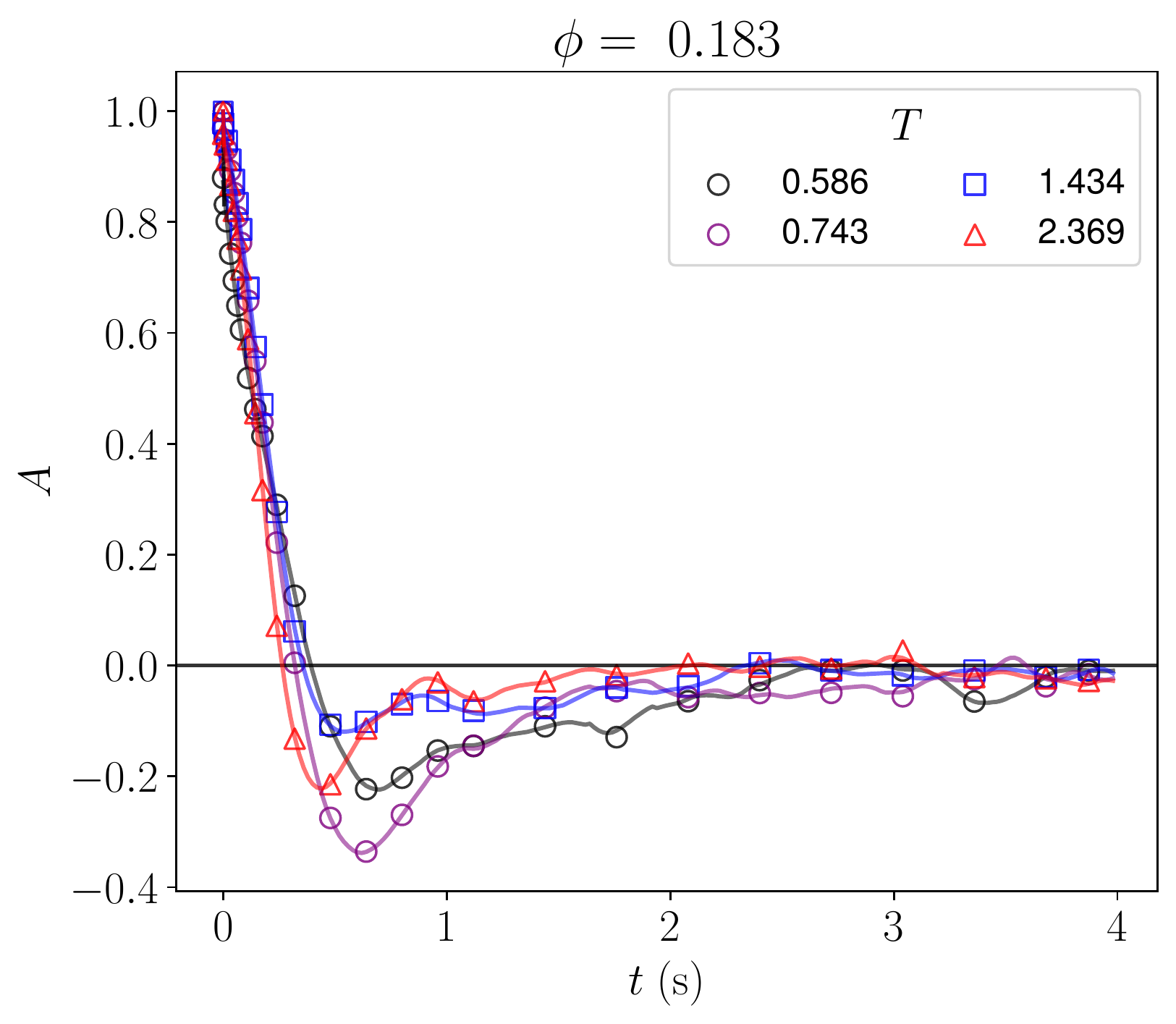} &
    \includegraphics[width= 0.3\textwidth]{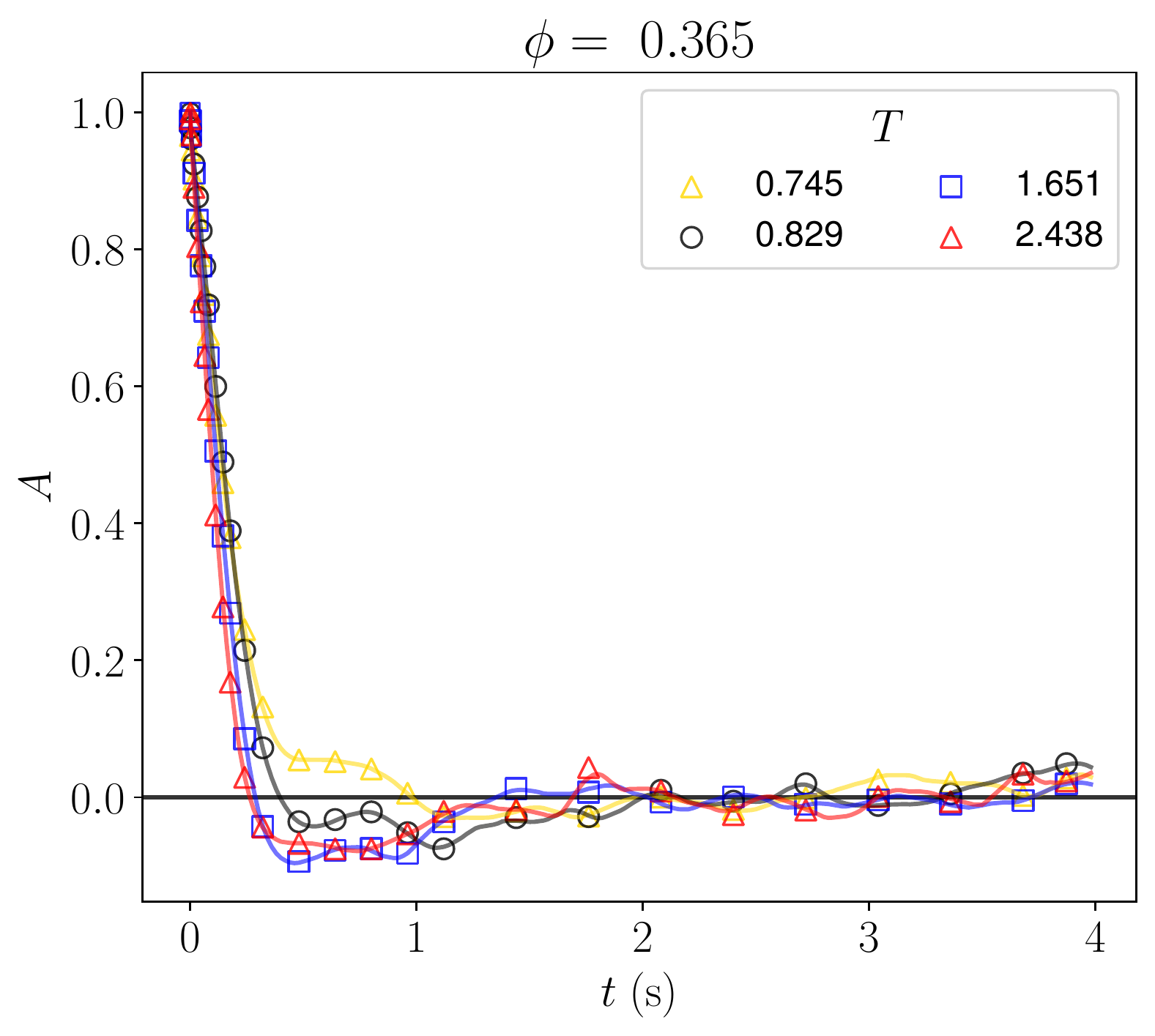} \\
    (c) \hspace*{0.25\textwidth} & (d) \hspace*{0.25\textwidth} \\
    \includegraphics[width= 0.3\textwidth]{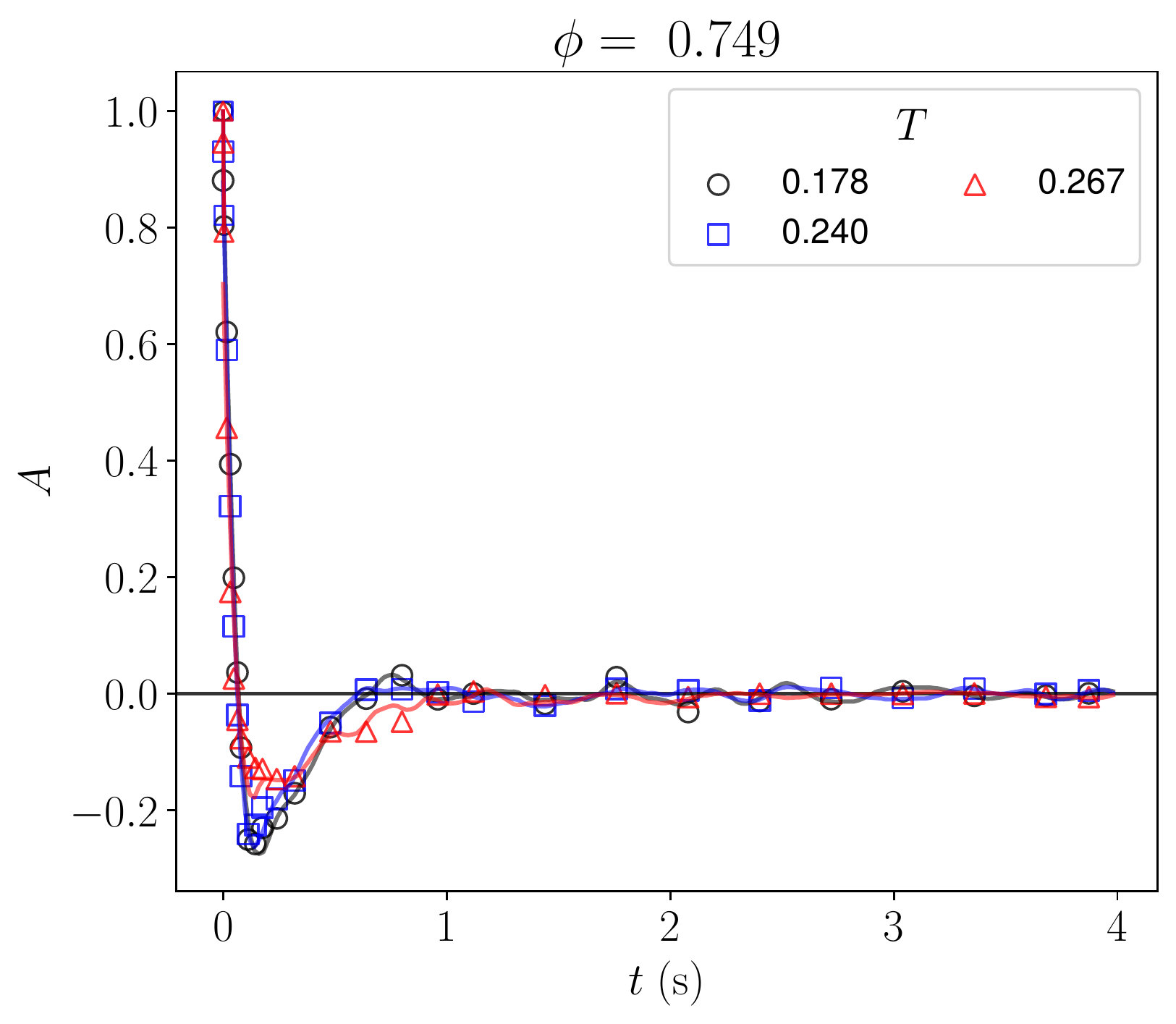} &
    \includegraphics[width= 0.3\textwidth]{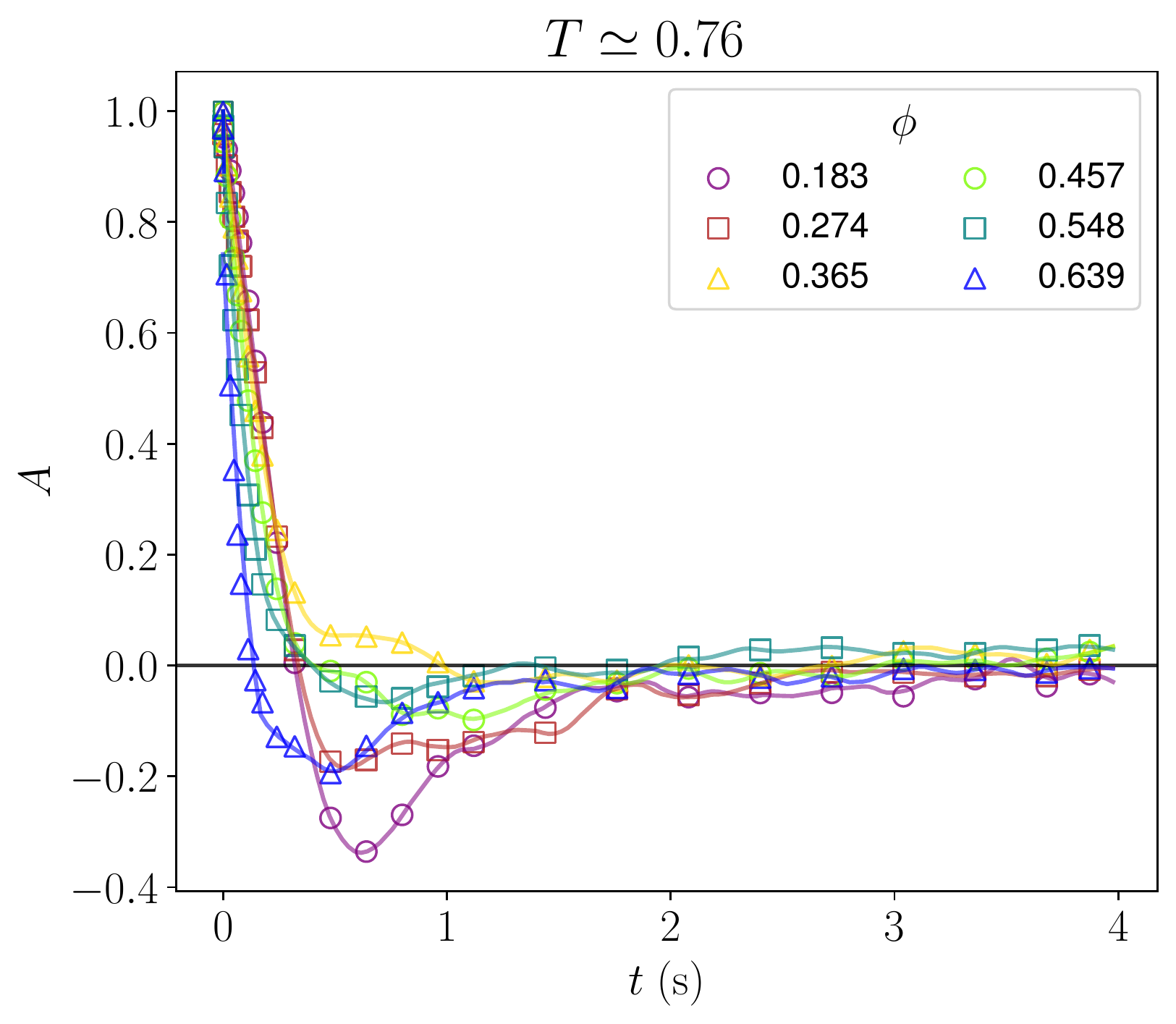}\\
  \end{tabular}
    \caption{Velocity autocorrelations. Panels (a), (b)
      \& (c) display data series taken at constant packing fraction ($\phi=0.18$, $\phi=0.365$
      \& $\phi=0.749$ respectively); (d) shows a data series taken at approximately constant
      granular temperature. }
    \label{vel_correlation}
  \end{figure*}

The velocity autocorrelation function (VAF) reflects the memory effects in the fluid and is related to key transport properties. Within our experimental accuracy, it has been verified to depend only on time differences. We thus define this quantity as follows:

\begin{equation}
A(t) = \frac{\langle \vec{v}(\tau)\cdot \vec{v}(t+\tau)\rangle}{\langle \vec{v}(\tau)\rangle^2},
\end{equation} where $\langle \dots \rangle$ indicate averaging over particles $i$ and time $t$, with a time step $\tau$.

Results are shown in Figure~\ref{vel_correlation}, where it can be readily noticed that there is a significant time
interval during which autocorrelations are negative. We interpret this as a clear indication of particle effects due to non-contact distance interactions mediated % -- after collision--
by the circulating air, %between neighboring particles
 as opposed to the behavior for hard
particles \cite{SGIW6}. %; i.e., the hydrodynamic interactions characteristic of particle suspensions.
Moreover, the decay time to negative autocorrelations can be regarded as a
measure of the typical collision time (in this context, ``collision'' should be
understood as a particle entering a region where it can feel the repulsive forces as it approaches other neighboring  particles). This collision time has been found to decrease with increasing density.

In order to characterize this effect,
Figure~\ref{vel_correlation} (d) presents measurements of the velocity autocorrelation for a wide range of densities at nearly constant temperature.  The displayed results clearly indicate that
non-contact interactions are in general more important at both ends of the density
spectrum. At very low densities the negative dip in the time behavior can be due to
a single-particle effect (e.g. vortex shedding). At lower densities the negative
values extend even up to $t\sim 2~\mathrm{s}$, indicating that the particles are
caged by their neighbors. Interestingly, the behavior is not monotonic, and in the
very dense regime, the dip becomes more pronounced again. This indicates that the
interstitial hydrodynamic effects are more complex than expected, this having an
impact in the phase behavior of the system, as results reveal later. Notice for
instance the curve for $\phi=0.365$, with only negative values at short times,
presents the behavior analogous to that of a gas, whereas for both lower and higher
densities stronger negative
autocorrelations at longer times show up, which is the behavior that can be expected for a
liquid. However, as diffusive properties will reveal, is at lowest density
($\phi=0.183$, purple symbols curve) where we
can actually detect the strongest negative autocorrelations, this indicating that what we are detecting
is actually a glass phase.  Finally, at very high densities, negative correlations
become stronger than in the liquid, this being a precursor evidence of symmetry
break (crystals developing). Thus, velocity autocorrelations seem to suggest the
following phase sequence for increasing density: glass, gas, liquid, crystal.

% \sby{Creo que debiera suprimirse la frase que sigue y la ecuación siguiente dado que esa ecuación no se usa y, además, si la difusión es anómala, no sirve para hallar el coeficiente difusivo anómalo}
% The time correlation function shows the dynamics of the fluid and the connection with the transport coefficients, self-diffusion, is given by the Green-Kubo formulae \cite{M54,R57}. The self-diffusion coefficients can be
% obtained as

% \begin{equation}
% D = \int_{0}^{\infty} \frac{1}{2}A(\tau) dt
% \end{equation}

\subsection{Diffusion}

%In our experiments, the balls move in a highly erratic way (see the movies in the Supplementary Material \cite{suppl}).
An important characteristic of the experiment particles random motion is the mean square displacement (MSD)  $\langle   r^2 \rangle$. Most frequently, systems exhibit a power-law long-time behavior of the MSD, i.e., $\langle  r^2 \rangle \sim D_\alpha t^\alpha$, where $\alpha$ is the \textit{diffusion exponent}, whereas $D_\alpha$ is the \textit{diffusion coefficient}. If $\alpha\neq 1$, the diffusion process is anomalous; in particular, it is called subdiffusive when $\alpha<1$.

\begin{figure}[t]
  \centering
  \includegraphics[width=.80 \columnwidth]{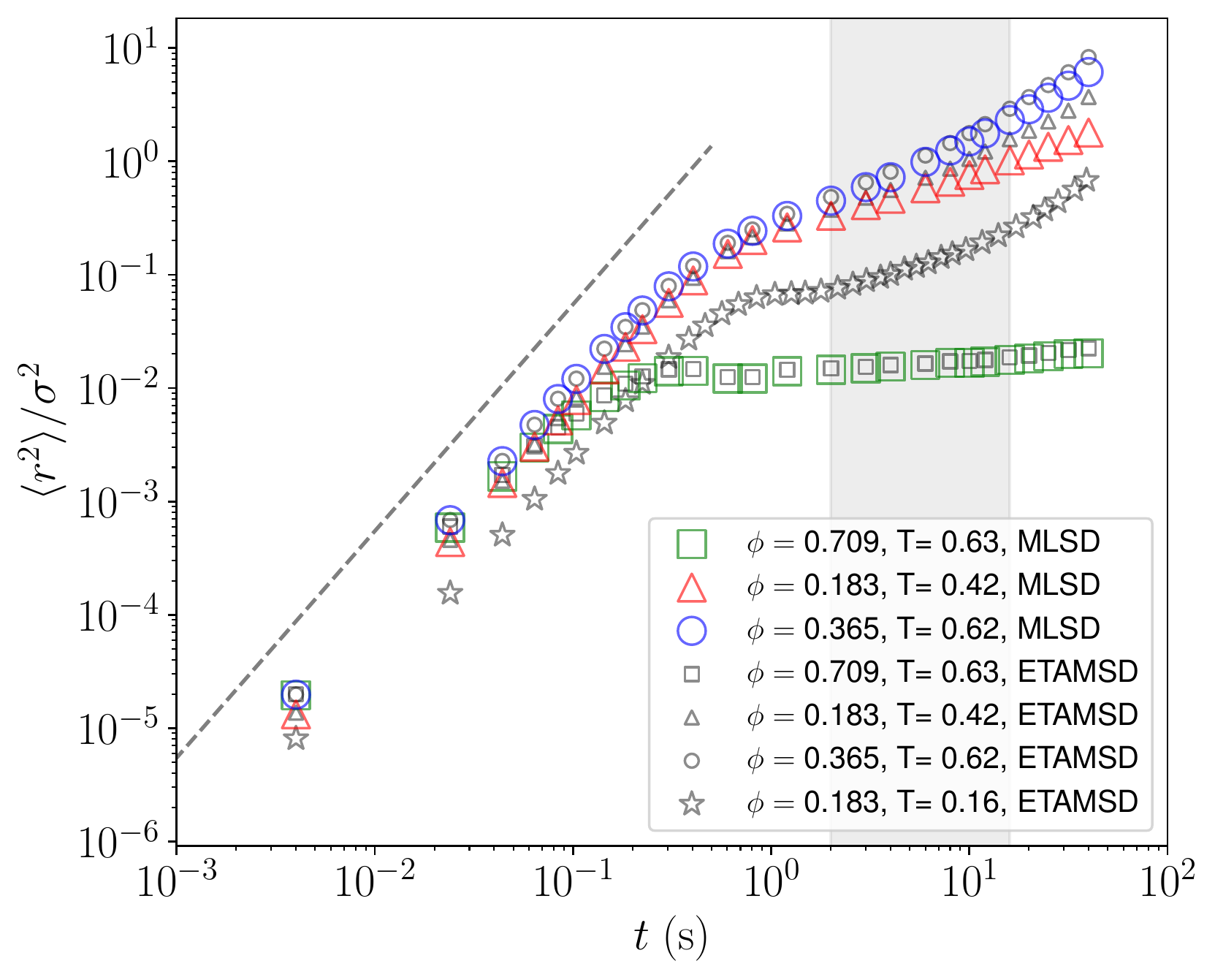}
  \caption{MLSD (large symbols) and ETAMSD (small symbols) vs. time for three experiments with $\phi=0.183$  and $T=0.422$ (triangles),   $\phi=0.365$   and $T=0.618$ (circles), and  $\phi=0.709$  and $T=0.632$ (squares). The dashed line has a slope equal to 2 characteristic of ballistic behavior. In the diffusive regime (gray region, corresponding to times between $t=2$ s and $t=16$ s) one has different slopes for different parameter sets. Star symbols correspond to a glassy transition, that typically displays a short plateau forming between ballistic and diffusive regimes.}
   \label{fig:MSD}
 \end{figure}

%In what follows, we will only be  concerned with the properties of $\alpha$. For normal diffusive processes one has %$\alpha=1$. When $\alpha\neq 1$, one distinguishes between two subcases respectively termed ``superdiffusion'' ($\alpha>1$), %and ``subdiffusion'' ($\alpha<1$).

%Evaluation of $\alpha$ in real systems is usually tricky \cite{Woringer2020}, since experimental verification of the power %law $\langle r^2 \rangle \propto t^\alpha$ requires (i) a large number of experiments (trajectories)  for an accurate %computation of the ensemble average and, (ii), a long enough time span over which the aforementioned power-law behavior %holds, so that the corresponding power-law fit is statistically meaningful. In many systems (and also in our setting), these %two requirements are not always guaranteed.

The drawback of only having at our disposal a limited number of trajectories can be alleviated through the standard procedure \cite{Meroz2015,Kepten2013} of constructing the time average of the mean square displacement (TAMSD) for each trajectory,
\begin{equation}
\label{x}
\overline{r^2(t)}=\frac{1}{t_m-t} \int_0^{t_m-t} d\tau \left|\vec r(\tau+t)-\vec r(\tau)\right|^2,
\end{equation}
($t_m$ is the measurement time) and subsequently taking the mean over the ensemble of time averages for the individual trajectories. This yields the ensemble average of the time averaged mean square displacement (ETAMSD)  $\langle \overline{r^2(t)} \rangle$.

In this procedure it is assumed that both the MSD $\langle  r^2 \rangle$ and the TAMSD  $\overline{r^2(t)}$ follow the same power-law dependence $t^\alpha$, so that $\alpha$ can be accurately computed from a limited number of trajectories. However, this is not always the case. A well-known counterexample exhibiting non-equivalence between the TAMSD and the MSD as a result of weak ergodicity breaking is transport generated by the so-called continuous-time random walk (CTRW) model \cite{Meroz2015}.
Fortunately, in our experimental system, there are no indications of such a behavior (for example, our VAFs are qualitatively different from those obtained from the CTRW model \cite{Burov2011}).

\begin{figure}[t]
  \centering
  \includegraphics[width=.80 \columnwidth]{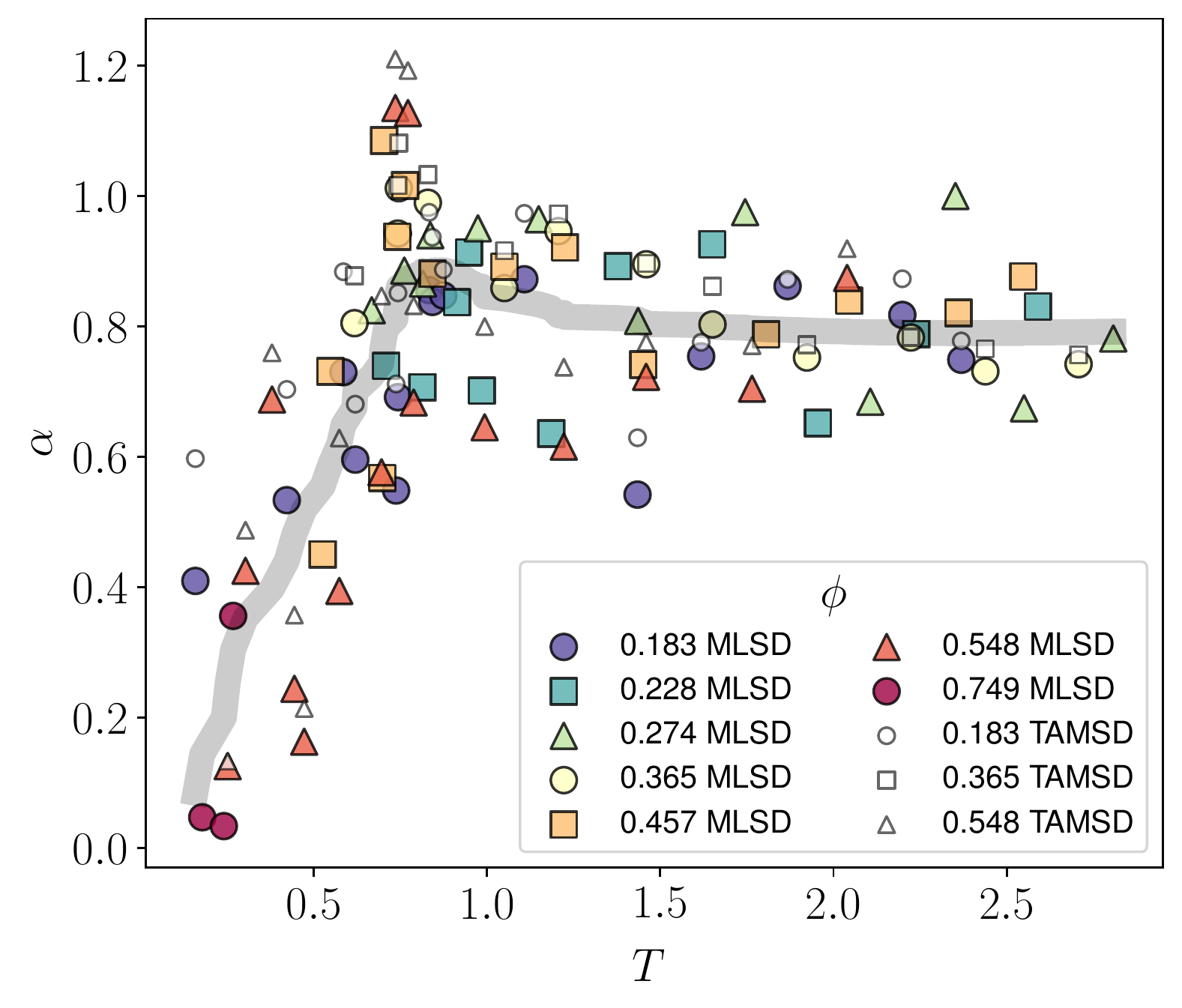}
  \caption{Diffusion exponent $\alpha$ vs. the temperature for several densities. As in Fig.\ref{fig:MSD}, the values of $\alpha$ were obtained by fitting the ETAMSD and MLSD curves between $t=2$ s and $t=16$ s. To generate such curves, only trajectories with a minimum length of 40 s have been taken into account. The thick solid gray line is a guide describing the general trend (a smoothing of the MLSD data points has been carried out with a $3^{\mathrm{rd}}$ order Savitzky-Golay moving polynomial).}
  \label{fig:alphavsT}
\end{figure}

%In our experiments, the limitation in the available number of useful trajectories, i.e., trajectories long enough to be used %for a reliable computation of the ETAMSD, mainly stems from two factors.  First, the number of balls is usually around a few %tens, in high-density systems at most two hundred; clearly, this number is not big enough for reliably carrying out ensemble %averages. Second, the \emph{effective} length of the random trajectories of the balls is relatively short, which admittedly %limits the quality of the estimate obtained for the TAMSD. More precisely, the duration of our longest recorded trajectories %is around 100 s, and the initial ballistic regime lasts around half a second (see below), which implies around 200 %independent random steps.  In addition, on many occasions, a ball is seen to exit the region of interest (ROI) while being %tracked, which implies an even shorter recorded trajectory.   Thus, in the best-case scenario, the ``clean'' random %trajectories that we have at our disposal for evaluating the ensemble average (i.e., trajectories not involving exits from %the ROI) consist of a few hundreds of elementary displacements only. This means that our estimates of $\alpha$ should be used %with caution, mainly for qualitative discussions.

In Figure~\ref{fig:MSD} we show some representative ETAMSD curves obtained from three experiments with $N=60$ ($\phi=0.183$, $T=0.422$), 120 ($\phi=0.365$, $T=0.618$) and 233 balls ($\phi=0.709$, $T=0.632$).
%% cases a, e, i   of figure \ref{hist2d}  %%%
 Only data corresponding to trajectories longer than 40 $s$ are considered. We have carried out fits of the EATMSD in the time interval $2\;\text{s}<t<16\; \text{s}$ (gray region in Figure~\ref{fig:MSD}). This choice is a trade-off ensuring that such an interval starts well after the end time of the ballistic regime, but is at the same time short enough to yield a sufficiently long time window $t_m-t$, so that statistical problems in the computation of the time average can be largely avoided.

In a further effort to obtain an improved estimate of $\alpha$,  we have also plotted curves displaying the time dependence of the so-called mean logarithmic square displacement (MLSD) \cite{Kepten2013}, which is the ensemble average of the logarithm of the TAMSD, $\log \overline{r^2(t)}$. A fit of this quantity as a function of $\log t$ leads, in general, to better estimates for $\alpha$, provided that the localization error in the particles position remains small (as is the case in our experiments) \cite{Kepten2013}.
%%%%

All curves clearly exhibit an initial ballistic regime during which $\langle\overline{r^2(t)}\rangle\sim t^2$. This holds up to times $\lesssim 0.1 s$. The ballistic regime is always followed by a subdiffusive regime ($\alpha<1.0$).
%%% COMPROBAR QUE ES CORRECTO in most of our experiments seems to be a transient regime that lasts until normal diffusion is eventually established.
For $\phi=0.183 $ and $\phi=0.365$ one can spot an increase in the slope of the final part of the experimental curves, which could indicate the eventual onset of normal diffusion at even longer times, not covered by our experiment. This terminal increase in the slope has indeed been found to be a typical feature in granular dynamics experiments (e.g., in ref. \cite{Abate2006}, both a transient subdiffusive regime and a final normal diffusion regime were identified for a proper parameter choice). Nevertheless, one should bear in mind that the quality of the TAMSD deteriorates for larger values of $t$, since for such values the size $t_m-t$ of the time window over which the average is performed decreases.
The offset of the MLSD and ETAMSD lines, quite noticeable for $\phi=0.183$, is not completely unexpected \cite{Kepten2013} . In this case the fit of the MLSD and ETAMSD curves between $t=2$ s and $t=16$ s leads to $\alpha=0.53$ and $\alpha=0.70$, respectively.  However, the case with $\phi=0.365$ leads to $\alpha=0.8$ and $\alpha=0.9$, respectively, whereas the case with $\phi=0.709$ leads to $\alpha=0.1$ for the two curves. The noteworthy difference in the values of $\alpha$ for $\phi=0.183$ turns out to be a persistent feature in our experiments, see Figure~\ref{fig:alphavsT}.

In Figure~\ref{fig:alphavsT}, we plot the values of $\alpha$ obtained by using the
time interval  $2\;\text{s}<t<16\; \text{s}$ to fit the MLSD. As a reference, we also
provide the values of $\alpha$ obtained from the ETAMSD computed for a number of
experiments. As in Kepten \emph{et al.} \cite{Kepten2013}, we have found that these $\alpha$-values
are generally higher than those yielded by the MLSD (the difference is around one or,
at most, two tenths); yet, they follow the same qualitative behavior.

The results in Figure~\ref{fig:alphavsT} reveal a large variability of $\alpha$, due statistical limitations inherent to our experiments (number of trajectories and limited movie clips duration due to camera memory limitations).
For example, for $\phi=0.365$ and $T=0.618$, the MLSD value of $\alpha$  shown in
Fig.~5 is $0.81$ (as already mentioned, this value follows from a fit in the
interval $2\;\text{s}<t<16\;\text{s}$ for  trajectories longer than
$40\;\text{s}$). However, if ones uses trajectories longer than $30\;\text{s}$, one
gets $\alpha=0.82$, and if one uses the interval $1\;\text{s}<t<20\;\text{s}$, one
finds $\alpha=0.80$. These three different values of $\alpha$ respectively become
equal to $0.90$, $0.83$ and $0.92$ if one chooses $T=1.461$, and for $T=1.651$ they
are $0.80$, $0.89$ and $0.77$.  These cases illustrate the kind of variability in
the value of $\alpha$ that we observe. In any case, if one changes the minimal
length of the trajectories and/or the fitting interval in a sensible way, one finds
that the corresponding values of $\alpha$ are compatible with the general
qualitative behavior shown in Fig.~5, and in this sense the latter is robust. We
note two main regimes, according to the behavior with respect to granular
temperature,  separated by a small region around $T=0.7$ (where the values of
$\alpha$ are close to 1, the normal diffusion exponent).  
At low temperatures ($T\lesssim 0.7$), $\alpha$ is clearly increasing with $T$.
We see that  $\alpha$   remains fairly small for the lowest measured granular
temperatures. In particular, we see that there are cases with  strong subdiffusive
behavior with $\alpha$ values well below $\alpha=0.5$. 
Interestingly, these are precisely the cases where the velocity distribution function deviates to a greater extent from a Maxwellian form (see  Figure~\ref{log_fv}). At higher temperatures, we find a second diffusive regime for which $\alpha$ displays a plateau vs $T$ (or at least, is not clearly decreasing nor increasing) and for which the values are still subdiffusive but noticeably larger  ($\alpha\sim 0.8 $) than at very low temperatures.

We think
that the strong subdiffusive behavior ($\alpha \le 0.5$) observed for sufficiently low temperatures
is likely due to cage effects \cite{Starr2013}. 
In fact, as density increases, one observes the onset  of a crystallization process; see Figures 7(g-h) in section \ref{phases}
 (crystals are typically
colder than granular fluids under the same forcing conditions \cite{LVU09}).  According
to \cite{Starr2013}, the cage size is identified as the value of $\langle r^2\rangle^{1/2}$ for which its logarithmic derivative $d \ln\left(\langle r^2\rangle^{1/2}\right)/d(\ln t)$ attains a minimum. It is interesting to note that in those cases where the ballistic behavior changes to strong subdiffusion (small $\alpha$), the cage effect is so strong that the MSD is even seen to decrease during a short crossover regime.  This effect can be clearly observed in the curves corresponding to  $\phi=0.709$.  We ascribe this behavior to the same transient viscoelastic forces that are responsible for the first dip exhibited by the VAF when $\phi=0.183$ and $\phi=0.709$ (cf. Figs.~\ref{vel_correlation}(a) and \ref{vel_correlation}(c)).
In fact, we  conjecture that the anomalous subdiffusive behavior found in non-crystalline phases is due to transient viscoelastic forces characteristic of complex interacting systems with correlated components \cite{Meroz2015}.

On the contrary, the high temperature diffusive regime should correspond to regions of the phase space where the dynamics is dominated by a fluid phase \cite{OLDLD04}. In summary, strong indications of a rich phase behavior in this system emerge out of its diffusive properties. We will address this issue in more detail in the next section.

\section{Structural properties}
\label{structural}

\begin{figure*}[t]
  \centering
  \begin{tabular}{ c c }
    (a) \hspace*{0.25\columnwidth} & (b) \hspace*{0.25\columnwidth}\\
    \includegraphics[width= 0.375\textwidth]{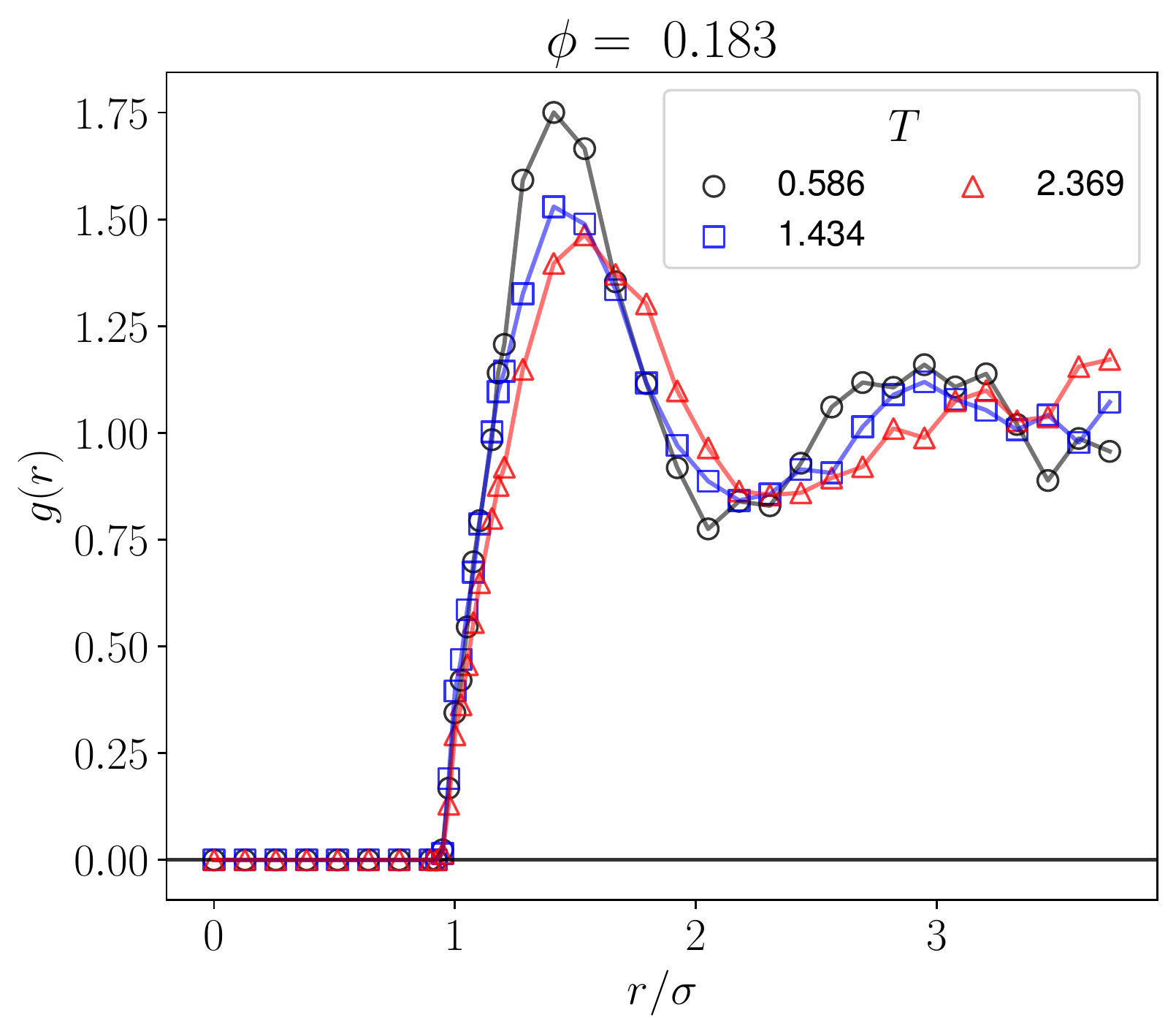} &
    \includegraphics[width= 0.375\textwidth]{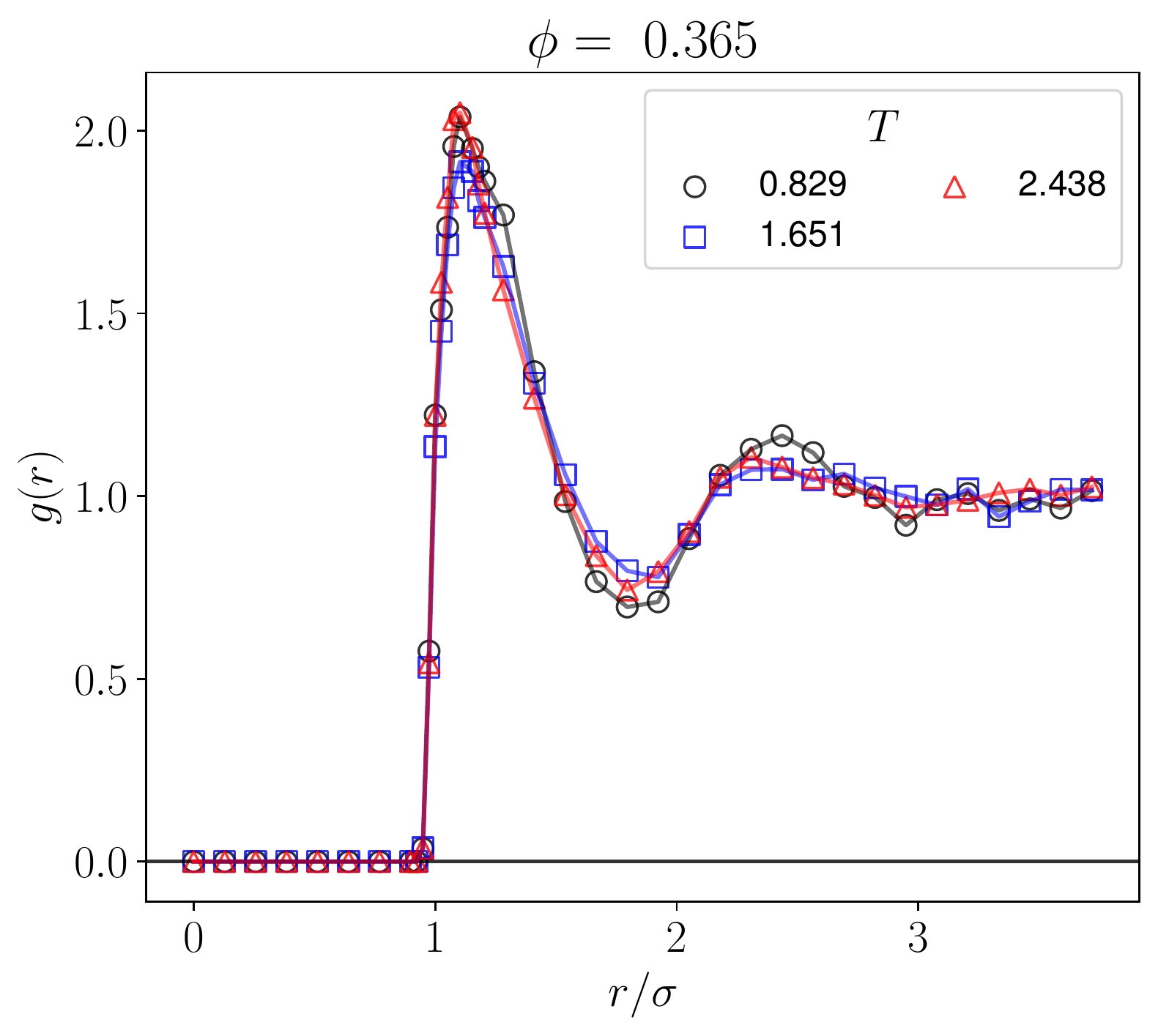} \\
    (c) \hspace*{0.25\columnwidth} & (d) \hspace*{0.25\columnwidth}\\
    \includegraphics[width= 0.375\textwidth]{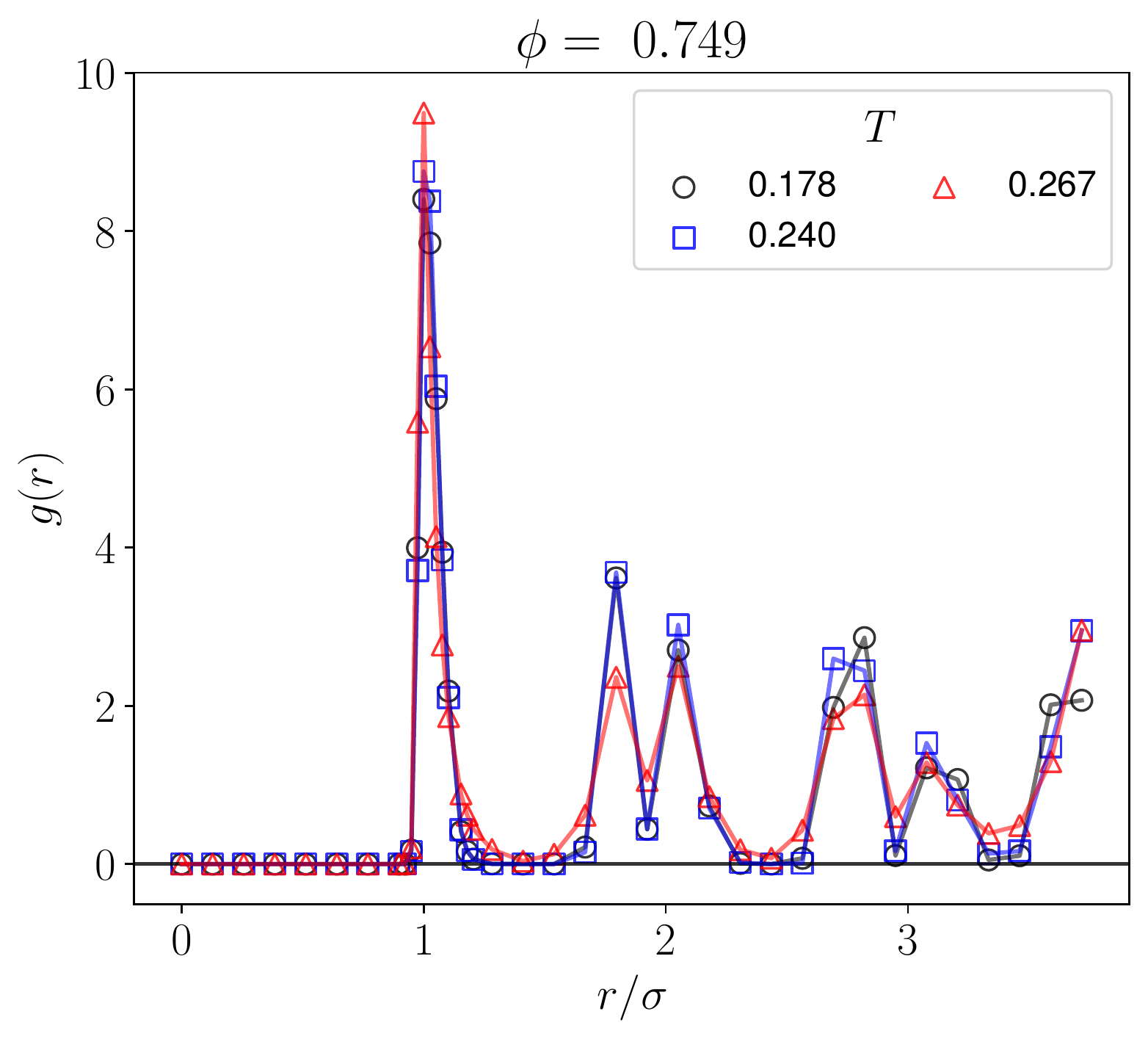} &
    \includegraphics[width= 0.375\textwidth]{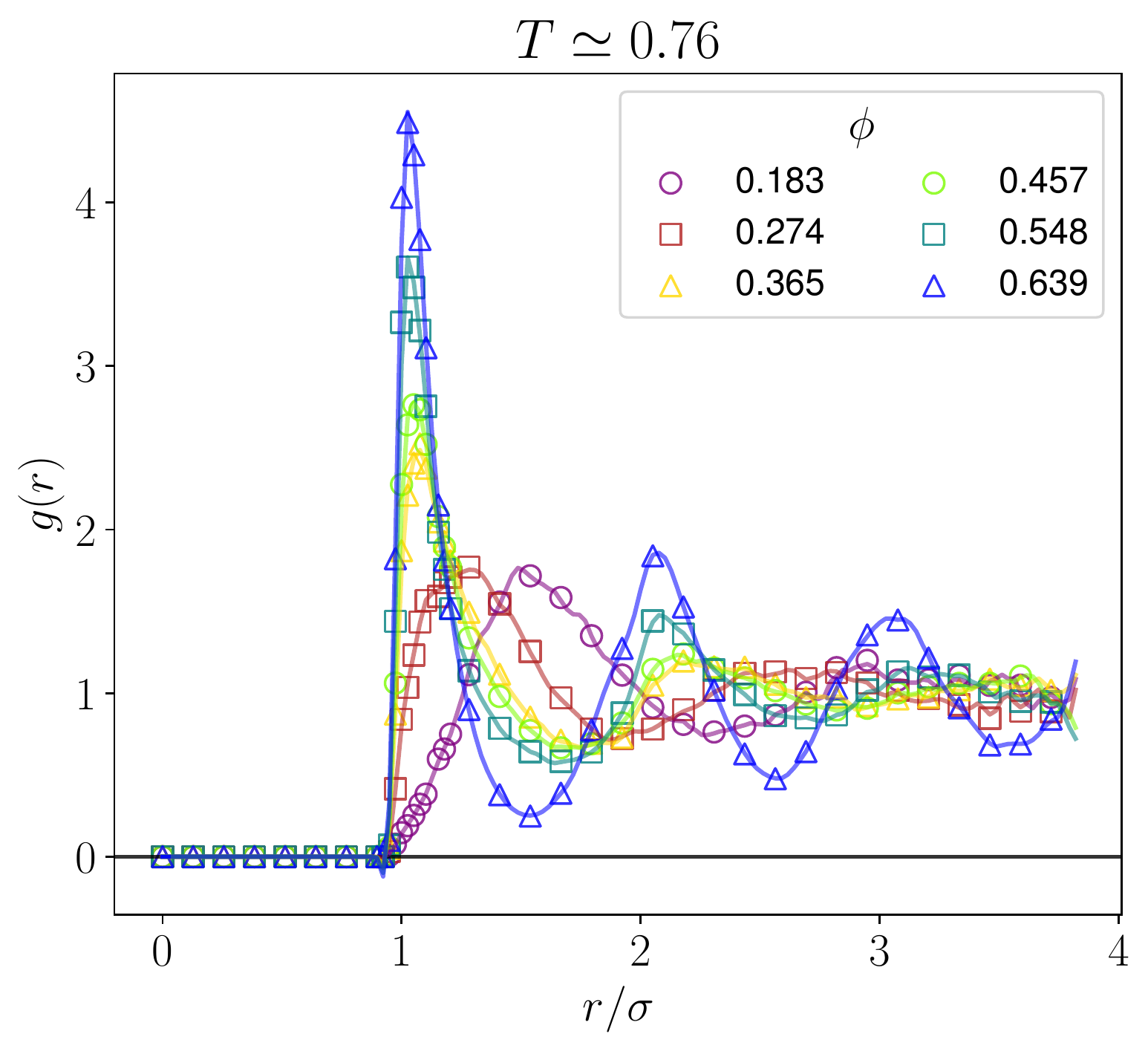}
    \end{tabular}
  \caption{Pair correlation function as processed from experimental $xy$
    particle positions. Panels (a), (b)
      \& (c) display data series taken at constant packing fraction ($\phi=0.18$, $\phi=0.365$
      \& $\phi=0.749$ respectively); (d) shows an data series taken at approximately constant
      granular temperature. A crystallization process is evident
    from the (a)-(c) panel series; clearly, complete crystallization is
    attained for $\phi=0.749$ (see cf.  panel (c)).}
  \label{gr_fig}
\end{figure*}

Phase transitions and crystallization processes were
analyzed in the 2D system \cite{Lemaitre1993,Ippolito1995,Oger1996} with disks but,
to our knowledge, not in the pseudo-2D system with spheres. Since interactions between spheres are mediated by strong long-ranged hydrodynamic forces, the phase
behavior can be expected to differ importantly from that of the system with disks,
where long-ranged forces have not been detected.

Thus, we devote this section to structural properties and phase
transitions. Motivated by the lack of previous data on phase transitions in this
system, we perform here a comprehensive analysis based on the pair correlation
function $g(r)$ and the Voronoi tessellation with the aim of uncovering as
thoroughly as possible the phase transitions landscape \cite{NR07}. As we will see, $g(r)$ already yields
 clear indications of different ordering transitions in the system. Voronoi
 tessellation is a graphical representation that partitions space in cells enclosing
 only one particle, so that all the points inside a given cell are closer to the
 associated particle than to any other particle in the system \cite{NR07}. This
 representation will confirm the expectations arising from the behavior of
 $g(r)$. Moreover, Voronoi tessellation also conveys additional structural
 information, thereby providing clear evidence for the onset of hexagonal order at
 high densities \cite{Lemaitre1993,OU05}.

\subsection{Radial distribution function}

Following a standard procedure, we have computed the radial distribution function
from our experimental data; taking into account that the system is 2D and has
constant particle density $\phi_0$, we employ the following formula: 

\begin{equation}
    g(r) = \sum_{i,j>i}^N\frac{1}{2\phi_0\pi r_{ij} \mathrm{d}r} \Pi(r_{ij}-r-dr/2)\,,
    \label{gr}
\end{equation}
where $\Pi(r_{ij}-r-dr/2)\equiv\Theta(r-r_{ij})\Theta(r+dr-r_{ij})$ is the
rectangular pulse function \cite{R12} ($\Theta$ being the Heaviside function \cite{nist}).

Measurements of the radial distribution function reveal interesting structural
changes in the system, as already advanced in the previous sections.  Results are
displayed in Figure~\ref{gr_fig}. As we can see, 
for $\phi=0.183$ (panel a) there is a liquid-like structure that is highly dependent on temperature. Notice that, in this case, the main peak appears at a distance clearly larger than $r=\sigma$ (we recall $\sigma$ is the particle diameter).
At a higher density ($\phi=0.365$, panel b) an analogous liquid-like structure emerges, but in this case it
is very robust against temperature variations. At even higher densities ($\phi=0.749$, panel c), we can
clearly see a series of sharp peaks, denoting positional ordering. These peaks have
been observed in previous studies \cite{Lemaitre1991} and their positions are related
to the reticular parameter in hexagonal packing. For instance, the secondary peak at
$r\lesssim2\sigma$ for instance corresponds to particles in two non-consecutive
vertexes in a hexagonal cell,  with one vertex in between, while the secondary peak at
$r\gtrsim2\sigma$ corresponds to particles in a hexagonal cell located at two non-consecutive vertexes and with two
intermediate vertexes. The pattern actually repeats around
$r\sim3\sigma$, out of neighbor hexagonal cells, thus indicating long-ranged spatial
correlations, inherent to a crystal.

Finally, in Figure~\ref{gr_fig}~(d), the behavior of $g(r)$ for different densities
is displayed in a series of curves at nearly constant temperature,  where we can clearly see the transition from fluid-like to crystal-like $g(r)$ curves as the density is increased. It is worth pointing out that sharp secondary
peaks already appear at densities as low (as compared to disks \cite{Lemaitre1991,Lemaitre1993}) as $\phi\sim0.6$, which is an indication of
a lattice parameter that is larger than the particles diameter. Furthermore, the
first secondary peak develops around $r=2\sigma$, this being a feature that appears
in a crystallization process. Note that this behavior is
reminiscent of that observed in early subcooled molecular liquids close to the glass
transition \cite{AJL19}. The pair correlation function reveals the emergence of some kind of spatial correlations and
 translational symmetry, but it does not provide information on the geometrical
 properties of this symmetry. For that purpose, the Voronoi diagrams, complemented
 with 2D histograms of particle positions, we present in
 the next subsection are more adequate.

 \subsection{Phase changes}
 \label{phases}

\begin{figure*}[htbp]
  \centering
  \begin{tabular}{c c c}
   (a) \hspace*{0.02\textwidth} $\phi=0.183$, $T=0.16$ \hspace*{0.02\textwidth} & (b) \hspace*{0.02\textwidth} $\phi=0.183$, $T=0.59$ \hspace*{0.02\textwidth} & (c) \hspace*{0.02\textwidth} $\phi=0.183$, $T=0.74$ \hspace*{0.02\textwidth} \\
    \includegraphics[width=0.32\textwidth]{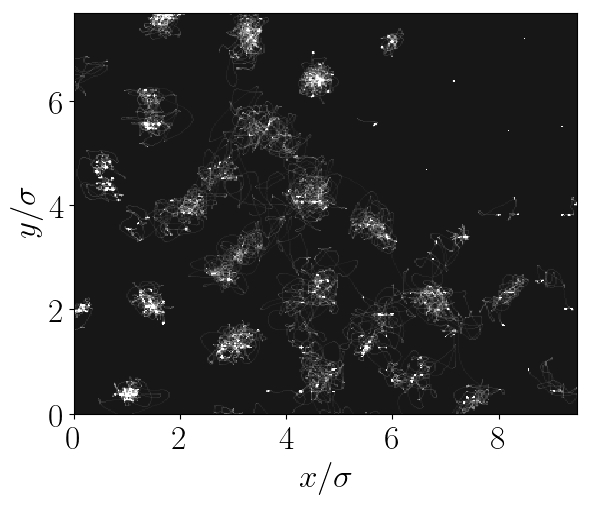}&
\includegraphics[width=0.32\textwidth]{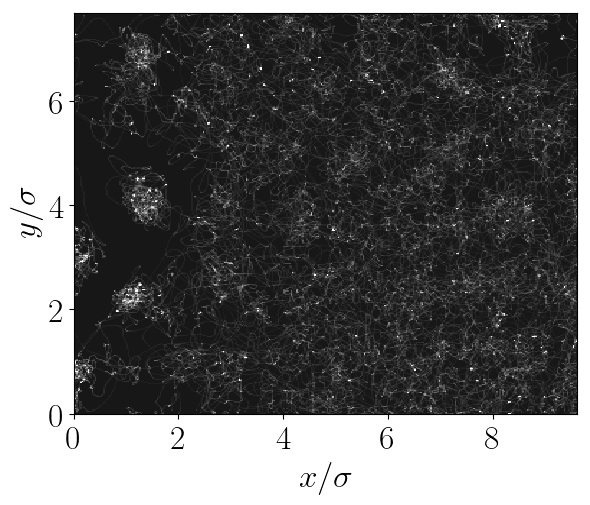}&                                                 \includegraphics[width=0.32\textwidth]{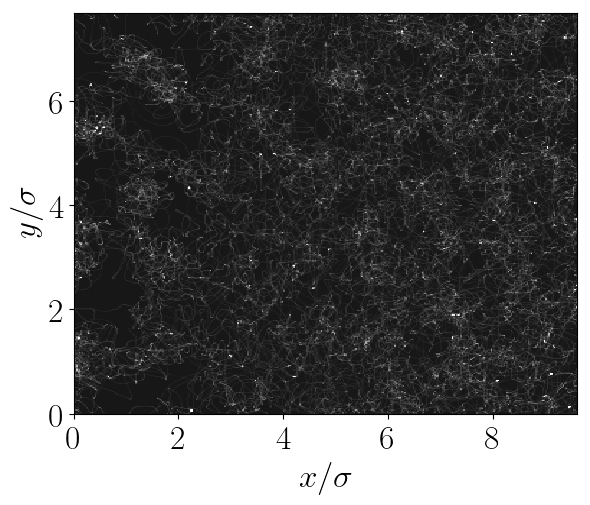}\\
   
   (d) \hspace*{0.02\textwidth} $\phi=0.274$, $T=0.67 $ \hspace*{0.02\textwidth} & (e) \hspace*{0.02\textwidth} $\phi=0.365$, $T=0.62 $ \hspace*{0.02\textwidth} & (f) \hspace*{0.02\textwidth} $\phi=0.457$, $T=0.70 $ \hspace*{0.02\textwidth} \\
    \includegraphics[width=0.32\textwidth]{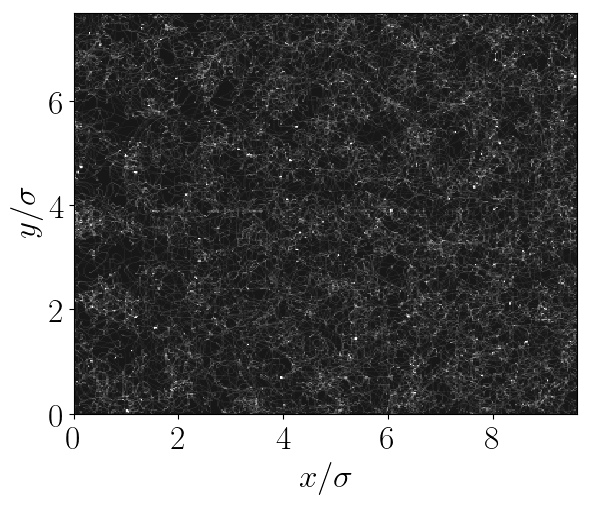}& \includegraphics[width=0.32\textwidth]{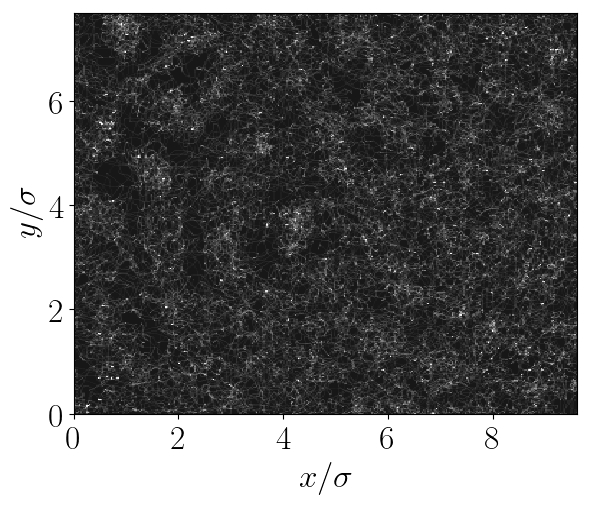}&
\includegraphics[width=0.32\textwidth]{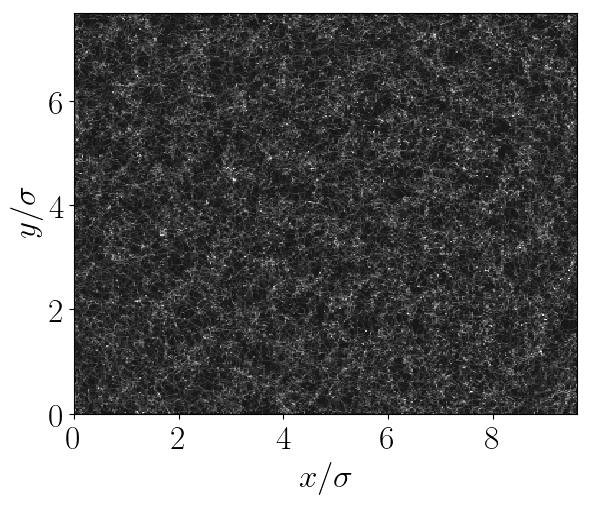}\\ 
   
   (g) \hspace*{0.02\textwidth} $\phi=0.548$, $T=0.70 $ \hspace*{0.02\textwidth} & (h) \hspace*{0.02\textwidth} $\phi=0.639$, $T=0.68 $ \hspace*{0.02\textwidth} & (i) \hspace*{0.02\textwidth} $\phi=0.709$, $T=0.63 $ \hspace*{0.02\textwidth} \\
 \includegraphics[width=0.32\textwidth]{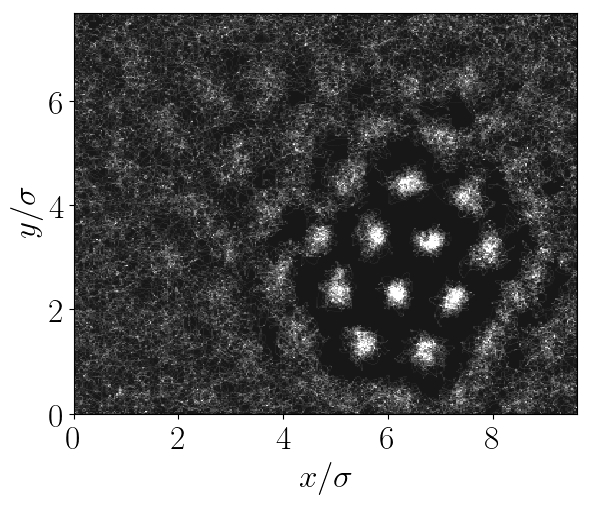}&
   \includegraphics[width=0.32\textwidth]{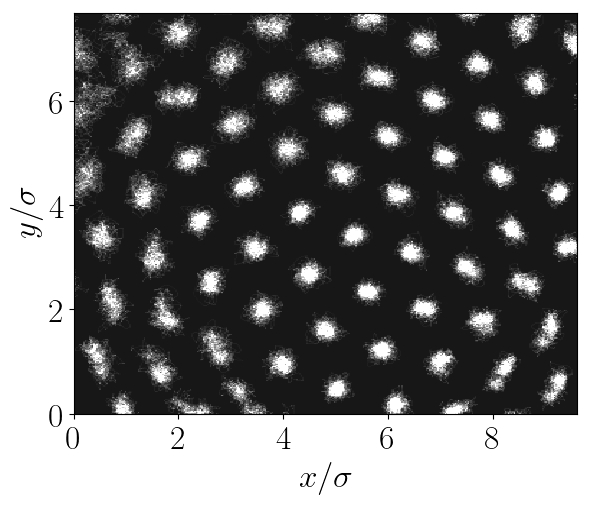}& \includegraphics[width=0.32\textwidth]{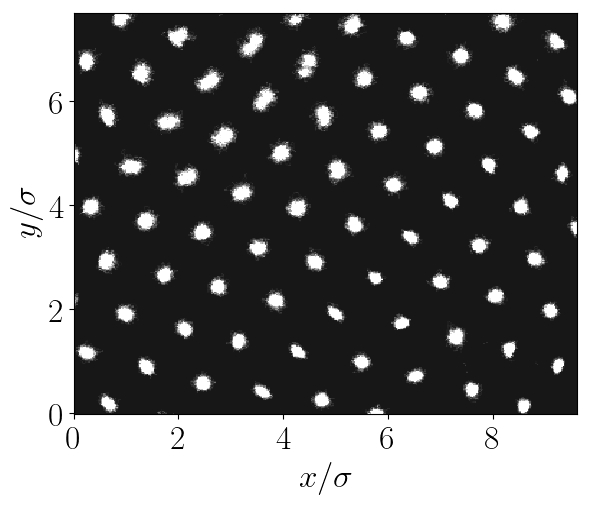}
  \end{tabular}
  \caption{Set of 2D histograms for different system configurations. These histograms have been generated from the complete set of images in the movie clips, each grey dot representing a particle's instantaneous position. First row: each figure corresponds, from left to right, to increasing granular temperature at constant density; as the system heats up, first glassy
    behavior, and then transition to liquid are observed. Second and third rows:
    each figure corresponds, from left to right, to increasing density at constant temperature; a transition from liquid to crystal
    takes place, with phase coexistence.}
  \label{hist2d}
\end{figure*}

\begin{figure*}[t]
  \centering
    \begin{tabular}{c c c}
   (a) \hspace*{0.02\textwidth} $\phi=0.183$, $T=0.16 $ \hspace*{0.02\textwidth} & (b) \hspace*{0.02\textwidth} $\phi=0.183$, $T=0.59 $ \hspace*{0.02\textwidth} & (c) \hspace*{0.02\textwidth} $\phi=0.183$, $T=0.74 $ \hspace*{0.02\textwidth} \\
    \includegraphics[width=0.32\textwidth]{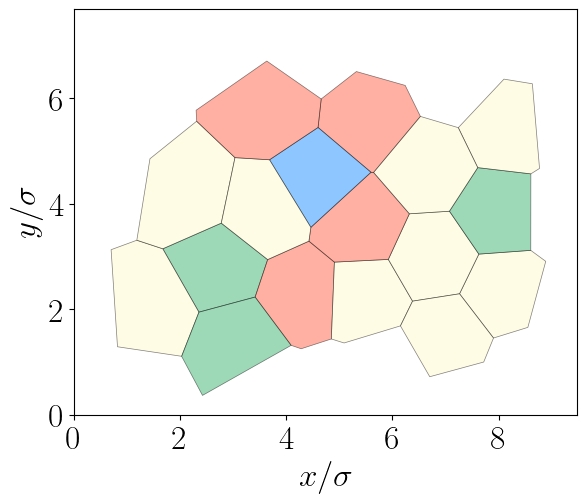}&
\includegraphics[width=0.32\textwidth]{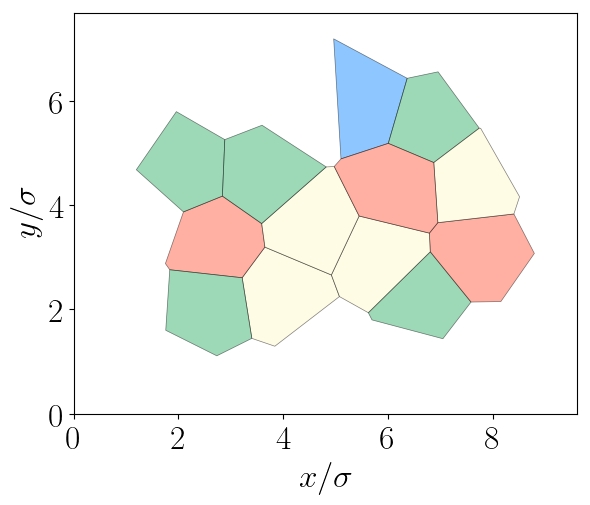}&                                                   \includegraphics[width=0.32\textwidth]{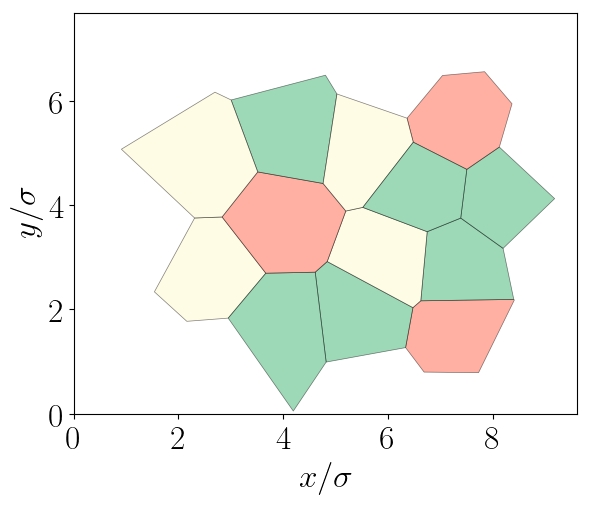}\\
   (d) \hspace*{0.02\textwidth} $\phi=0.274$, $T=0.67 $ \hspace*{0.02\textwidth} & (e) \hspace*{0.02\textwidth} $\phi=0.365$, $T=0.62 $ \hspace*{0.02\textwidth} & (f) \hspace*{0.02\textwidth} $\phi=0.457$, $T=0.70 $ \hspace*{0.02\textwidth} \\
    \includegraphics[width=0.32\textwidth]{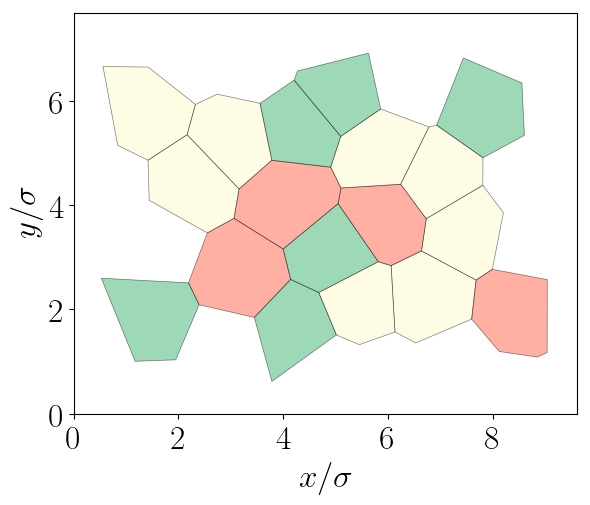}& \includegraphics[width=0.32\textwidth]{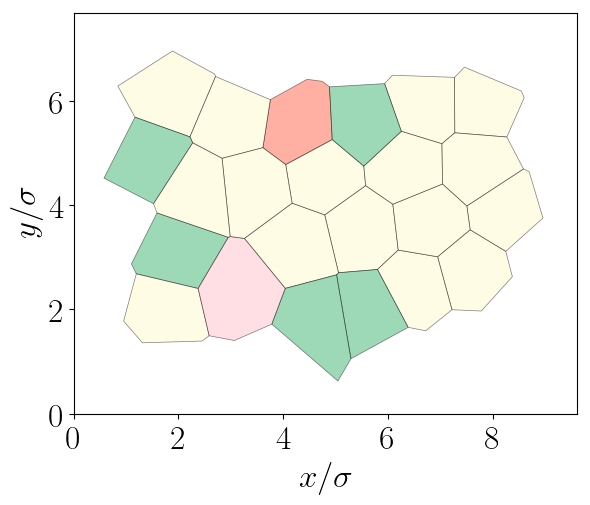}&
\includegraphics[width=0.32\textwidth]{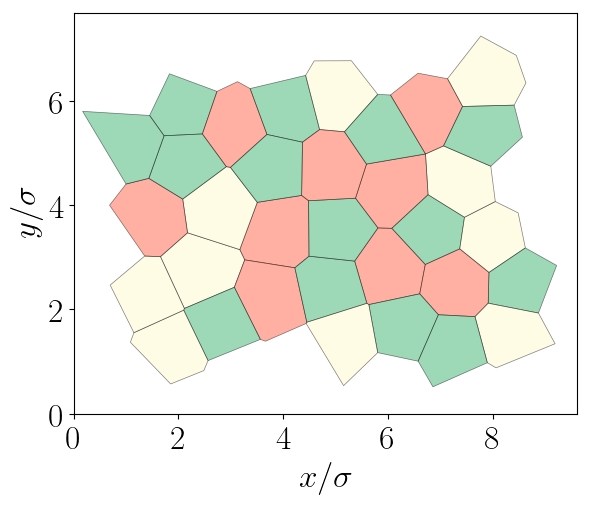}\\ 
   (g) \hspace*{0.02\textwidth} $\phi=0.548$, $T=0.70 $ \hspace*{0.02\textwidth} & (h) \hspace*{0.02\textwidth} $\phi=0.64$, $T=0.68 $ \hspace*{0.02\textwidth} & (i) \hspace*{0.02\textwidth} $\phi=0.709$, $T=0.63 $ \hspace*{0.02\textwidth} \\
 \includegraphics[width=0.30\textwidth]{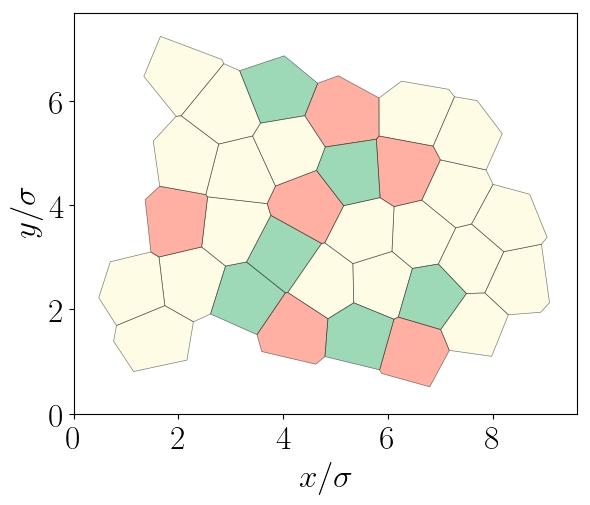}&
   \includegraphics[width=0.30\textwidth]{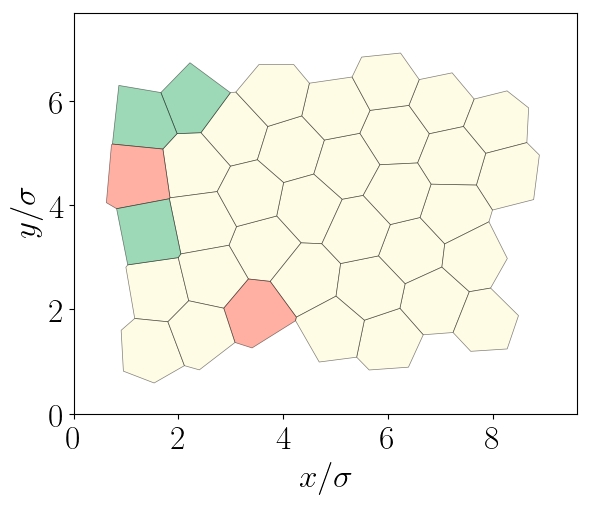}& \includegraphics[width=0.30\textwidth]{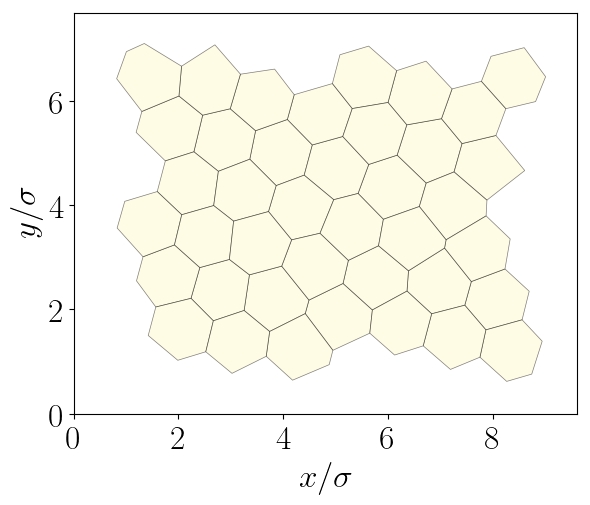}
  \end{tabular}
  \caption{Voronoi diagrams corresponding to the systems depicted in
    Figure~\ref{hist2d}. These diagrams confirm the glassy behavior in
    panels (a) and (b), the lack of order in the liquid-like systems, see panels
    (d), (e), (f), and the emergence (initially in coexistence with a liquid
    phase) of hexagonal ordering, see panels (g), (h). (i). (Irregular) polygons
    have been marked according to the following color code: blue for squares, green for pentagons, yellow for hexagons, dark red for heptagons, and light red for octagons.}
  \label{voronoi}
\end{figure*}

In order to detect emerging structural changes, we explored large regions of the
parameter space (see supplementary material \cite{suppl} for a list of experimental data). Results below clarify that performing an exhaustive set of experiments at different densities (and granular temperatures) was necessary since the phase behavior is very rich and complex, which otherwise would have remained unnoticed. For instance, the
series with varying particle density at constant temperature shows a very rich and
peculiar phase behavior. In order to visualize the varying degree of symmetry and the
qualitative changes in related properties, we use both 2D spatial histograms and
Voronoi tessellation diagrams \cite{NR07}, see Figures~\ref{hist2d} and
~\ref{voronoi}.

As already anticipated in Section~\ref{ExpM}), each of the 2D histograms depicted in Figure~\ref{hist2d} visualizes the positions of each particle (represented by white pixels) averaged over the time duration of each movie  ($\sim 100~\mathrm{s})$, as usual in previous studies on phase transitions \cite{PMKW78,OU98}. In contrast, Figure~\ref{voronoi} represents
Voronoi diagrams \cite{NR07} of instantaneous states of the same system. Both figures complement each other, i.e., the 2D spatial histograms tell us about the persistence in time of a given geometrical structure, whereas the Voronoi tessellation diagrams inform us about the specific geometry of that
structure.

The evidence provided by Figures~\ref{hist2d},~\ref{voronoi} is rather compelling in spite of the fact that diagrams at very low
densities ($\phi < 0.1$), for which the dynamics is primarily driven by individual particles (particle-particle interactions are not frequent, and the
dynamics is expected to be very similar to what has been previously reported for analogous single-particle systems \cite{OLDLD04}). In contrast, it is interesting to note that for somewhat larger densities $ 0.15 \lesssim \phi \lesssim 0.25$, we observe
glass-like states at sufficiently low temperatures. These glassy phase are
characterized by particles staying trapped by their neighbors (cage effect) for a
sufficiently long time until they can escape to another cage where again then they
remain during a long time again and so on. This is reflected in the diffusive
behavior shown in Figure~\ref{fig:MSD} in the stars symbols series ($\phi=0.183,
T=0.16$), displaying a characteristic plateau (i.e., there is an intermediate
region, here at $t\sim1$, for which the curve is horizontal) in the MSD
\cite{Rodriguez-Rivas2019}.  Indeed,  Figure~\ref{hist2d}~(a) reveals significant  
inhomogeneities in particle dynamics, with sections of the system (bottom left and
top right) where particles 
positions are more persistent in time; on the other hand, the corresponding Voronoi 
tessellation (Figure~\ref{voronoi}~a) shows a variety of cells with different  coordination numbers, which signals the absence of a clearly dominant symmetry structure. At higher densities ($ 0.25 \lesssim \phi \lesssim 0.5$), only a single phase is observed,
which is seemingly disordered and isotropic, and can therefore be regarded as
liquid-like [see panels (c)--(e)]. In panel (f), we have noted an even higher degree
of disorder, with particles distributed in a more uniform fashion. Interestingly enough, upon further increase of
the density ($\phi \gtrsim 0.5$), we see the development of areas of hexagonal
ordering in coexistence with the fluid phase [see panels (g,h)]. The hexagonally ordered
phase grows with increasing density -- panel (h)-- eventually occupying
the entire system, see panel (i).

\begin{figure}[ht]
  \centering
  (a)\newline\includegraphics[width=0.45\textwidth]{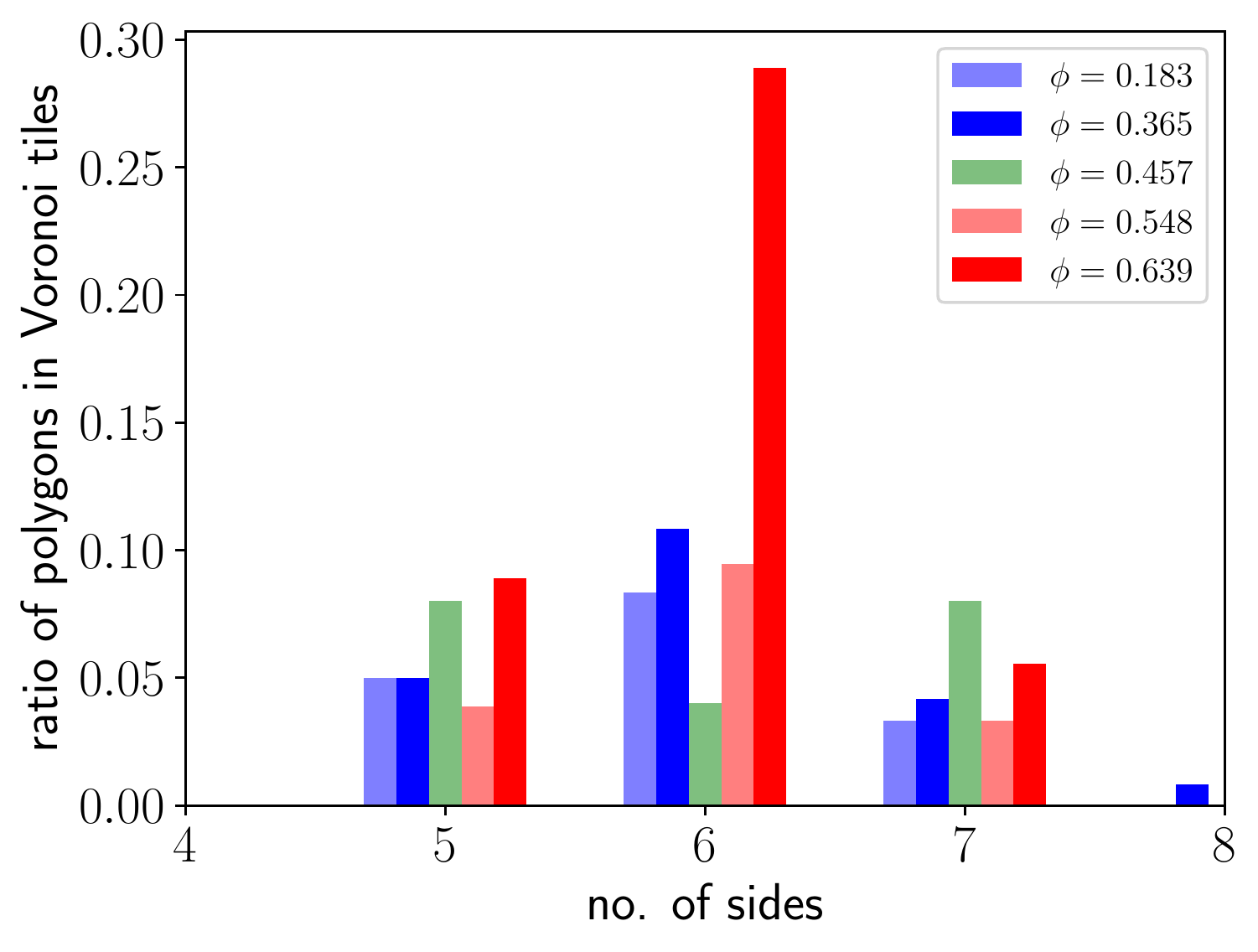}\\
  (b)\newline\includegraphics[width=0.45\textwidth]{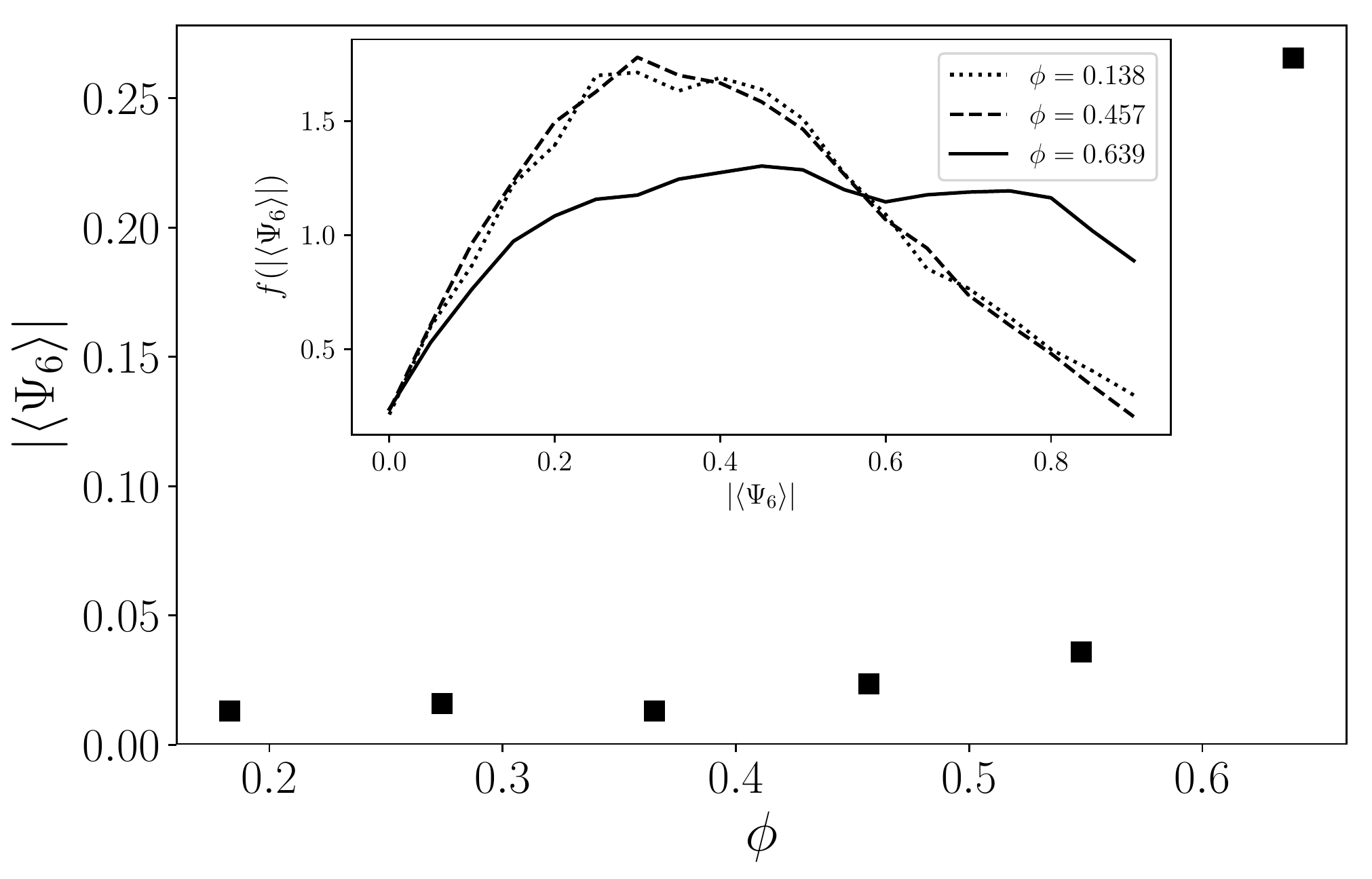}
  \caption{(a) Voronoi tiles histograms, displaying time averaged distributions for different densities. (b) Average of six-bond order parameter,x
    $|\langle\Psi_6\rangle|$,  for constant temperature series
    ($T\simeq0.76 $). Its density distribution function is
    also represented in the inset, for the cases $\phi=0.183$ (glass, pointed line),
    $\phi=0.457$ (liquid, dashed line) and $\phi=0.639$ (hexagonal crystal,
    continuous line).\label{fig:psi6} } 
\end{figure}

It is interesting to note also that hexagonal ordering appears at much lower
densities ($\phi \simeq 0.5$ and higher) than in systems of hard particles \cite{OU05}, in experimental assemblies of disks \cite{Lemaitre1991} or in soft disk
models used in molecular dynamics to mimic active or passive particles, see
e.g. \cite{DLCGP19}). We think this signals strong effects of long-ranged
hydrodynamic forces between the rolling spheres over the phase
behavior. Furthermore, we clearly detected phase coexistence of the hexagonal
crystal with the liquid phase
(Figure~\ref{hist2d} g). This notably differs from the observations of
hexagonal crystallization/melting in a confined monolayer of vertically
vibrated and quasi-elastic spheres \cite{OU05} and in air-fluidized disks
\cite{Lemaitre1993}. In fact, in these two latter
systems the hexagonal crystal undergoes a melting transition of
KTHNY \cite{KT73,Halperin1978,NH79,Y79} type (i.e.; the melting transition for the
vibrated layer is continuous and mediated by the successive unbinding of dislocation
and disclination pairs, and it does not involve phase coexistence). Finally, it is
also remarkable that for very cold systems there is 
no collective ordered collapse as it happens in a vibrated monolayer of hard
spheres \cite{OU98}. Instead, the particles in the gas phase gradually and undergo
unstructured strong
freezing, which results in the collective formation of disordered lattices (Figure 7
a). 

In addition to Figures~\ref{hist2d},~\ref{voronoi}, movie clips and experimental data
of the different phases observed are available in the Supplementary
material \cite{suppl}. The complexity and richness of the phase transitions that
we have observed is worth being studied in more detail. Such a study will be carried
out in subsequent works.

Finally, in Figure~\ref{fig:psi6} (a) we represent in a more quantitative manner
geometrical configurations for glass, liquid and hexagonal crystal, by means of
Voronoi histograms (averaged over all frames) according to the particle coordination number (or, equivalently, type
of polygon for each Voronoi tile). As we can see, hexagonal cells become
predominant only when the hexagonal crystal is fully developed (for $\phi\ge0.6$), whereas in both glass and liquid the cell distribution is more
uniform. In order to quantify the liquid-hexagonal more specifically, we represent in
Figure~\ref{fig:psi6}(b) the absolute value of the average of the six-bond order
parameter $\Psi_6$ for an increasing density series, at constant temperature, and
its density distribution functions for three different densities. This order
parameter average is defined, for each frame, as
$\Psi_6=(1/N)\sum_k^N(1/N_k)\sum_j\mathrm{e}^{6\theta_{ik}\i}$, where $\theta_{jk}$ is the
bond angle for the $k-j$ particle pair and the $j$ sum runs over the $N_k$ neighbors of
particle $k$ (the sum over the $k$ particle index is the magnitude averaging, assuming the
system has $N$ particles in total). After this, we average for all frames, which we
denote as $\langle\Psi_6\rangle$. A steep increase in $|\langle\Psi_6\rangle|$ is noticeable for
packing fraction $\phi>0.548$, which is the density corresponding to the system with
position 2D histogram in Figure~7 (g), for which we first
find a developing hexagonal crystallite. It is also to be noticed here that glass and
liquid present rather similar behavior; i.e. cell histograms and six-bond order
parameter do not display ordering, which is what we expected since these phases
are indistinguishable off their structural properties.

\section{Conclusions}
\label{conclusions}

We have studied in this work the pseudo-2D dynamics of a set of air-driven identical
spheres which, excited by turbulent air, roll under Brownian movement on a
horizontal metallic grid.

To the best of our knowledge, we have obtained the first experimental series showing
the influence of particle density on the behavior of the distribution function
(Figures~\ref{log_fv} d) at nearly constant 
temperature. The distribution function exhibits non-Maxwellian high energy tails, a feature also
 reported in previous works on granular dynamics \cite{OU98,NE98}. However, in contrast with
 the behavior of a monolayer of vertically vibrated particles \cite{OU99}, these
 non-Maxwellian tails seem to be more prominent  at higher temperatures (Figures~\ref{log_fv} a, b, c).

% We have also characterized the exponential tails emerging in the high energy, which turn out to be more prominent in the high density regime. In the opposite limit of a dilute system, the
% distribution function clearly tends clearly to a Maxwellian; as a matter of fact, in the joint limit of
% low density and low temperature, the kurtosis of the distribution function does not display measurable deviations from 3, which is the value that corresponds to a Gaussian distribution. This agrees
% with previous experimental results for the single particle dynamics of a similar system \cite{OLDLD04}.

Velocity autocorrelations illustrate the relevance of hydrodynamic forces due to
airflow-mediated  particle
interactions. Our analysis unveils an important difference with respect to analogous
experimental set-ups, like thin layers of vertically vibrated spheres \cite{OU05} or
air-fluidized disks \cite{Lemaitre1993}. In particular, we show that hydrodynamic
forces result in the onset large negative autocorrelations at comparatively short
times (Figure~\ref{vel_correlation}). Direct observation confirms that particles
initially approaching each other then experience an effective repulsion at distances.
This yields  a very peculiar phase
map, as is shown in section \ref{phases}. In particular, it prevents the formation of
gas-like  states at very low
densities ($\phi\lesssim 0.15$ ); contrary to the case of air-fluidized disks, we
observe independent Brownian-like behavior for each particle, which rarely
collide (see
Supplementary  material experiments clips \cite{suppl}).
Clearly, the effect of repulsive forces in the dynamics of the system is more
important at low densities, which is consistent with the observation that negative
autocorrelations at short times  are more pronounced in dilute
systems (Figure~\ref{vel_correlation}(d), see also movies in Supplementary
material \cite{suppl}).

Two distinct diffusive regimes have been observed. In contrast with previous works,
the system can remain subdiffusive even in disordered low density phases (Figure~\ref{fig:alphavsT}), which is another
consequence of the existence of long-ranged hydrodynamic interactions. We thus see no strong dependence of the diffusion exponent $\alpha$ on the particle density. In contrast, $\alpha$ turns out to be very sensitive to changes in the granular temperature. At very low temperatures, $\alpha$ takes very small values, and the system is strongly subdiffusive. At somewhat higher temperatures, the system still remains strongly subdiffusive despite the steady growth of $\alpha$ with increasing temperature. Finally, at temperatures $T\approx 0.7$ and higher, $\alpha$ stabilizes around values that are weakly subdiffusive and a plateau is observed.

% i.e., there
% is no clear trend for the diffusion vs. particle density. We have
% observed that the diffusive regime strongly depends on granular temperature
% instead. The first regime, at low temperatures, is strongly increasing with granular
% temperature. At the lowest temperature values the system is strongly
% subdiffusive. At slightly higher temperatures, diffusion approaches normal
% diffusion. In this second regime, at high temperatures, we reach a plateau with only slightly subdiffusive behavior.

The phase transitions observed in our system display a surprisingly rich and peculiar
behavior, not reported previously in similar systems and ranging from a collection of independent Brownian-like particles at low
densities to glassy or  liquid states at moderate densities, and to the onset of regular hexagonal
lattices at higher densities. Most notably, the hexagonal crystal
melting occurs here in coexistence with a liquid phase. This finding  differs
strikingly 
from previous results reported for 2D systems of air-fluidized 
disks \cite{Lemaitre1993} and quasi-2D systems of quasi-elastic spheres \cite{OU05}, where phase coexistence of
liquid and hexagonal crystals in the melting transition was not found. In our
system, the behavior of the liquid to
hexagonal crystal transition appears to be more similar to what has been reported for highly inelastic
spheres in a quasi-2D system \cite{KT15}, where phase coexistence has also been
reported. However, in our system the phase coexistence seems to be mediated by long-ranged
hydrodynamic forces rather than by the inelasticity of particle collisions and
because of this reason occurs at noticeably lower densities. Thus, further study on
the evolution of the bond-orientational 
correlation function \cite{Pasupalak2020} or the $p_n$ parameter distribution
\cite{RRFM14,MMBL03} will be needed to cast light on the precise mechanism of this
phase transition in future work.

While there is an extensive bibliography referring to engineering
applications of air table systems \cite{MBF81}, here we have used one such system for
a more fundamental purpose, namely, to describe a variety of non-equilibrium quasi-2D phase
transitions and to identify the analogies with and departures from equilibrium
theories and previous observations in granular dynamics experiments with
air-fluidized disks \cite{Lemaitre1991,Lemaitre1993} and thin vibrated layers \cite{Melby2005,KT15}.

Summarizing, our results unveil a very rich and original behavior of our quasi-2D
system at various levels (distribution function, diffusion, velocity and spatial correlations,
phase transition diagrams, etc.) with respect to its closest analogs. Furthermore, contrary to
first observations in quasi-2D granular systems \cite{OU05}, our results suggest that the hexagonal crystal melting
transition in granular systems may in general not follow the KTHNY scenario.

% Results convey a
% very original behavior (for distribution function, velocity auto-correlations and
% phase transitions, as we explaiend) of this kind of system with respect to its
% closest analogs.

\section*{Acknowledgements}
  The authors thank Dr. F. Rietz and Dr.  A. Lasanta for fruitful discussion and Prof. J. S. Urbach for valuable
  discussion and revision of the manuscript. We acknowledge funding from  the Government of Spain through project No. FIS2016-76359-P and from  the regional Extremadura Government through projects No. GR18079 \& IB16087, both partially funded by the ERDF.
%%%END OF MAIN TEXT%%%


\balance


%%%REFERENCES%%%
\bibliography{ppp1} %the name of your .bib file

\end{document}